%% file: ms.tex
\pdfoutput=1

\documentclass{elsarticle}

\input{macros}

\usepackage{flushend}

\newcommand{\KEYWORDS}{
Concurrency topics, 
concurrency topic hierarchy,
concurrency topic difficulty,
concurrency topic popularity,
concurrency topic sentiment,
data race tool usage,
\SO
}

\begin{document}

\title{
Interests, Difficulties, Sentiments, and Tool Usages of Concurrency
Developers: \\ A Large-Scale Study on \SO
}

\author[add1]{Mehdi Bagherzadeh\corref{cor}}
\ead{mbagherzadeh@oakland.edu}

\author[add1]{Syed Ahmed}
\ead{sfahmed@oakland.edu}

\author[add1]{Srilakshmi Sripathi}
\ead{ssripathi@oakland.edu}

\author[add2]{Raffi Khatchadourian}
\ead{raffi.khatchadourian@hunter.cuny.edu}

\cortext[cor]{Corresponding author.}

\address[add1]{Department of Computer Science and Engineering, Oakland University, MI, USA} 
\address[add2]{Department of Computer Science, City University of New York, Hunter College, NY, USA}
%
%
%
%

\urlstyle{sf}
 
\input{abstract}

\maketitle  

\input{introduction}
\input{methodology-new}

\input{results}

\input{discussion}

\input{validity}
\input{related}

\input{conclusion}

\section*{Acknowledgements}
Bagherzadeh, Ahmed, and Sripathi were supported in part by the Department of Computer Science and Engineering and the Research Office at 
Oakland University. 
Khatchadourian was supported by PSC-CUNY Awards \#617930049 and
\#638010051, jointly funded by The Professional Staff Congress and The City University of New York.
We would also like to thank Nicholas Fireman for his help.

\clearpage
\newpage
\bibliographystyle{elsarticle-num}
\bibliography{ref-full,bagherzadeh,khatchadourian} 

\mycomment{
\clearpage
\newpage

\input{letter}

\clearpage
\newpage
\input{structured-abstract}
}

\end{document}

%% file: macros.tex
\usepackage[utf8]{inputenc}
\usepackage{pifont}

\usepackage{listings}
\usepackage{enumitem}

\usepackage{dashrule}
\usepackage{graphicx}
\usepackage{color}
\usepackage{xspace}
\usepackage{url}
\usepackage{comment}
\usepackage[hidelinks]{hyperref}
\usepackage[skins]{tcolorbox}
\usepackage{tabularx}
\usepackage{colortbl}
\usepackage[mathscr]{eucal}

\usepackage{float}
\usepackage[caption = false]{subfig}

\usepackage{mathpartir}

\usepackage{multirow}
\usepackage{amstext}
\usepackage{amsmath}
\usepackage{natbib}

\usepackage{makecell}

\newcommand\para[1]{{\bf \textit{#1   }}}
\newcommand{\bi}[1]{{\bf \textit{#1}}} 
\newcommand{\nobi}[1]{#1} 
\newcommand{\Number}[1]{\emph{\small (#1)}}

\newcommand{\figref}[1]{Figure~\ref{#1}}
\newcommand{\tabref}[1]{Table~\ref{#1}}
\newcommand{\secref}[1]{Section~\ref{#1}}

\newcommand{\MehdiNoteDone}[1]{}

\newcommand{\SO}{Stack Overflow\xspace}
\newcommand{\Mallet}{MALLET\xspace}
\newcommand{\etal}{et al.\xspace}

\newcommand{\percent}[1]{#1\%}
\newcommand{\ppercent}[1]{(#1\%)}

\newcommand{\finding}[2]
{
\begin{center}
\begin{tcolorbox}[
     enhanced, sharp corners,
     width=\columnwidth,center upper,
    drop fuzzy shadow northeast,
    halign=justify,
    colframe=black,colback=white,
    left=2pt,right=2pt,top=2pt,bottom=2pt]
\bi{Finding #1:} #2 
\end{tcolorbox}
\end{center}
}

\newcommand{\canset}{\mathscr T}

\newcommand{\tagset}{\mathscr T}

\newcommand{\cpostset}{\mathscr C}
\newcommand{\ipostset}{\mathscr P}

\newcommand{\sopostset}{\mathscr S}

\newcommand{\kset}{\mathscr R}
\newcommand{\uset}{\mathscr U}

\newcommand{\msrspace}{}

\newcommand{\plusplus}[1]{\stepcounter{#1}\arabic{#1}}
\newcommand{\merge}[2]{#1}

\newcommand{\topic}[1]{\emph{#1}}
\newcommand{\otopic}[1]{\emph{#1}}
\newcommand{\vquestion}[2]{(#1):\emph{``#2''}}
\newcommand{\equestion}[2]{}
\newcommand{\squestion}[2]{\emph{``#2''}}

\newcommand{\category}[1]{#1}

\newcommand{\colorit}[2]{{\color{#1}#2}}
\newcommand{\hierarchy}[2]{#1$\backslash\newline$#2}
\newcommand{\shierarchy}[2]{#2}

\newcommand{\paradigm}{\category{programming paradigms}}
\newcommand{\model}{\category{concurrency models}}
\newcommand{\threading}{\category{multithreading}}
\newcommand{\processing}{\category{multiprocessing}}
\newcommand{\correctness}{\category{correctness}}
\newcommand{\persistence}{\category{persistence}}
\newcommand{\performance}{\category{performance}}

\newcommand{\Q}[1]{$\mathscr Q$.\emph{#1}}

\newcommand{\negspace}{}

\newcommand{\captionskipneg}{}
\newcommand{\captionskiptableneg}{}

\newcommand{\sig}{\mathscr \mu}
\newcommand{\rela}{\mathscr \nu}
\newcommand{\subscript}[1]{\emph{\text{#1}}}

\newcommand{\btag}[1]{\emph{#1}}
\newcommand{\SentiSD}{Senti4SD\xspace}

\newcommand{\signed}[3]{#1 / #3}
\newcommand{\coincidence}[1]{}
\newcommand{\configexec}{configuration \& execution}

\newcommand{\GitHub}{GitHub\xspace}

\newcommand{\ONumber}[1]{\ding{#1}}
\newcommand{\llstinline}[1]{\lstinline|#1|}

\renewenvironment{quote}%
  {\list{}{\leftmargin=0.1cm\rightmargin=0.1cm}\item[]}%
  {\endlist}
\newcommand{\bigquoted}[3]{\vspace{-2mm}\begin{quote}\Q{#1} \bi{#2} \emph{#3}\end{quote}\vspace{-2mm}}

\newcommand{\mycomment}[1]{}

%% file: abstract.tex
\newcounter{abscounter}
\setcounter{abscounter}{0}

\begin{abstract}
\para{Context}
Software developers are increasingly required to write concurrent code that is
correct.
However, they find correct concurrent programming challenging.
To help these developers, it is necessary to understand concurrency topics they
are interested in, their difficulty in finding answers for questions in these topics, their
sentiment for these topics, and how they use concurrency tools and techniques
to guarantee correctness.
Interests, difficulties, sentiments, and tool usages of concurrency developers
can affect their productivity.
\\\para{Objective}
In this work, we conduct a large-scale study on the entirety of \SO to
understand interests, difficulties, sentiments, and tool usages of concurrency
developers.
%
\\\para{Method}
To conduct this study, we take the following major steps. 
First, we develop a set of concurrency tags to
extract concurrency questions and answers from \SO.
Second, we 
group these questions and answers into concurrency topics, categories, and a
topic hierarchy.
Third, we  
analyze popularities, difficulties, and sentiments of these concurrency topics
and their correlations.
Fourth, we develop a set of race tool keywords to extract concurrency questions
about data race tools 
and group these questions into race tool topics. 
We focus on data races because they are among the most prevalent concurrency
bugs. 
%
Finally, we 
discuss the implications of our findings for the practice, research, and
education of concurrent software development, investigate the relation of our findings
with the findings of the previous work, and present a set of example questions that developers
ask for each of our concurrency and tool topics as well as categories.
\\\para{Results}
A few findings of our study are:
\ONumber{182}
questions that concurrency developers ask can be grouped into a hierarchy with 27 concurrency topics under 8 major
categories,  
\ONumber{183}
\topic{thread safety} is among the most popular concurrency topics and \topic{client-server concurrency} is among the least popular,
\ONumber{184}
\topic{irreproducible behavior} is among the most difficult topics 
and \topic{memory consistency} is among the least difficult,
\ONumber{185}
\topic{data scraping} is among the most positive concurrency topics and \topic{irreproducible behavior} is among the most negative,
\ONumber{186}
\topic{root cause identification} has the most number of questions for usage of data race tools and \topic{alternative use} has the least.
While some of our findings agree with those of previous work, others sharply contrast. 
\\\para{Conclusion}
The results of our study can not only help concurrency developers but also concurrency educators and researchers
to better decide where to focus their efforts, by trading off one
concurrency topic against another.
  
%
\end{abstract}

%% file: introduction.tex
\section{Introduction}
Software developers are increasingly required to write concurrent code to satisfy
not only functional but also nonfunctional requirements of their software.
For example, software with a graphical user interface must be 
concurrent to satisfy a functional requirement that it should be able to display more than one window at a time. 
Similarly, software with timing constraints must be concurrent to be able to satisfy a nonfunctional requirement that 
it should provide a better performance.    
Software is concurrent if its 
computations can  potentially  run at the same time and otherwise is sequential.
In addition to satisfying its functional and nonfunctional requirements, the
concurrent software must be correct. For example, with the initial value of
\lstinline|x = 0|, the software with the code \lstinline|x = x + 1|, 
that is executed by two concurrent threads, should set the value of
\llstinline{x} to \llstinline{2}, which is the correct and desirable value.
However, a concurrency bug, such as a data race bug, can set the value of
\llstinline{x} to the incorrect and undesirable value \llstinline{1}. 
This data race bug happens when both threads first read
the initial value \llstinline{0} of \llstinline{x} and then increment and set it to
the value \llstinline{1} separately.  
A data race happens when two concurrent computations access
the same variable of the program and at least one of these accesses writes to
the variable.
%
%
%
%
%
However, most developers think sequentially and find
writing concurrent programs that are correct 
challenging \cite{Lu:08,Khatchadourian:19}.

To help concurrency developers with writing concurrent code that is correct, it
is necessary to understand concurrency topics they are interested in, 
their difficulty in finding answers for questions in these topics, 
their sentiment for these topics, and how they use concurrency tools and
techniques to guarantee correctness.
Interests, difficulties, sentiments, and tool usages of concurrency developers
can affect their productivity \cite{Ortu:15,Souza:17,Garcia:13,DeChoudhury:13}.
This understanding can not only help concurrency developers and their practice but also the research and education of concurrent software development by allowing members of these communities to better decide when and where 
to focus their efforts 
\cite{Barua:14,Bajaj:14,Treude:11,Allamanis:13,Rosen:16,Yang:16,Neuhaus:10,Pinto:14}.
Without such understanding, practitioners may not prepare for similar difficulties, 
researchers may make incorrect assumptions about interests  of practitioners, and educators may teach the wrong concurrency topics.

With more than 3 million developer participants and 38 million questions and
answers, written in 2 billion words, 
\SO \cite{Stackoverflow} has
become a large and popular 
knowledge repository for developers to ask questions, receive answers, 
and  learn about a broad range
of topics.
This makes \SO a great source to
learn about interests, difficulties, sentiments, and tool usages of concurrency
developers
\cite{Bagherzadeh:20,Bagherzadeh:19,Barua:14,Bajaj:14,Rosen:16,Yang:16,Pinto:14,Pinto:15}.

In this work, we conduct a large-scale study on the entirety of \SO to understand 
 interests, difficulties, sentiments, and tool usages of concurrency developers
 by answering the following research questions:
\newcounter{rqcounter}
\setcounter{rqcounter}{0}
\begin{itemize}[nosep]
  \item \bi{RQ\plusplus{rqcounter}. Concurrency topics} What concurrency topics do  developers ask questions about?
  \item \bi{RQ\plusplus{rqcounter}. Concurrency topic hierarchy}   %
  What categories do these concurrency topics belong to?
  \item \bi{RQ\plusplus{rqcounter}. Concurrency topic popularity} What topics are more popular among concurrency developers?
  \item \bi{RQ\plusplus{rqcounter}. Concurrency topic difficulty} What topics are more difficult to find answers to their questions?
  \item \bi{RQ\plusplus{rqcounter}. Concurrency topic sentiment}  What topics are more sentimental among concurrency developers?
  \item \bi{RQ\plusplus{rqcounter}. Correlations of topic popularity, difficulty \& sentiment} 
  Are there correlations between topic popularity, difficulty, and sentiment?
  \item \bi{RQ\plusplus{rqcounter}. Concurrency tool topics}
  What topics do developers ask  questions about when using data race tools and techniques?
\end{itemize}
For tool usage, we focus on the tools for finding and
fixing data race bugs. We focus on data races because data race bugs 
are the not only one of the most prevalent bugs in concurrent
software  \cite{Lu:08,Bagherzadeh:20} but also
notoriously difficult to find and fix where some of these bugs can have extreme
consequences.
In a recent emperical study of concurrency bugs,
data race bugs are about half of the concurrency bugs \cite{Asadollah:17} in the
open source software. Data races were the culprits in the Therac-25 disaster
\cite{Leveson:93}, the Northeastern electricity blackout of 2003 \cite{Blackout}, and the mismatched NASDAQ Facebook share prices of 2012 \cite{Pozniansky:07}.
Therefore, numerous tools and techniques \cite{Asadollah:17} have been developed 
to help concurrency developers with data races. 

%
 

To answer our research questions, we take the following major steps. 
First, we develop a set of concurrency
tags to  extract concurrency questions and answers from \SO.  
Second, we use topic modeling \cite{Blei:03} and card sorting \cite{Fincher:05} techniques to group these questions and answers into concurrency topics. 
Third, we use card sorting \cite{Fincher:05} to group similar topics into
concurrency categories and construct a topic hierarchy.
Fourth, we use well-known 
metrics that are proposed and used by the previous work
\cite{Rosen:16,Yang:16,Bajaj:14,Treude:11,Rosen:16,Nadi:16,Barua:14}
to measure the popularities, difficulties, and sentiments of these topics  and analyze their correlations.
Fifth, we develop a set of race tool keywords to extract concurrency questions about data race tools and techniques. 
Sixth, we use card sorting to group these questions into race tool topics.   
Finally, we discuss the implications of our findings for the practice, research, and education of concurrent software development. 
We also investigate the relation of our findings with the findings of
the previous work and present a a set of examples of questions that developers
ask for each of our concurrency and tools topics and categories.

\newcounter{itemcounter}
\setcounter{itemcounter}{0}
A few findings of our study are: 
\\
\para{Concurrency topics} 
\Number{\plusplus{itemcounter}} 
Concurrency questions that developers ask  
can be grouped into 
27 concurrency topics  
ranging from \topic{multithreading} to \topic{parallel computing}, 
\topic{mobile concurrency} to \topic{web concurrency}, and \topic{memory consistency} to \topic{runtime speedup}.
\Number{\plusplus{itemcounter}} Developers ask more about the \topic{correctness} of their concurrent software than their \topic{performance}.
\\
\para{Concurrency topic hierarchy}
\Number{\plusplus{itemcounter}} 
Concurrency questions can be grouped into a hierarchy with 
8 major categories: 
\topic{concurrency models}, \topic{programming paradigms}, \topic{correctness},
\topic{debugging}, \topic{basic concepts}, \topic{persistence}, \topic{performance}, and \topic{graphical user interface (GUI)}.
\\
\para{Concurrency topic popularity}
\Number{\plusplus{itemcounter}} 
\topic{Thread safety} is among the most popular concurrency topics and \topic{client-server concurrency} is among the least  popular.
\\
\para{Concurrency topic difficulty}
\Number{\plusplus{itemcounter}} 
\topic{Irreproducible behavior} is among the most difficult concurrency topics 
and \topic{memory consistency} is among the least difficult.  
%
\\
\para{Concurrency topic sentiment}
\Number{\plusplus{itemcounter}} 
Most questions of all concurrency topics are neutral in sentiment.  
\Number{\plusplus{itemcounter}} 
\topic{Data scraping} is among the most positive concurrency topics and 
\topic{process life cycle management} is among the least positive.
\Number{\plusplus{itemcounter}} 
\topic{irreproducible behavior} is among the most negative concurrency topics and 
\topic{process life cycle management} is among the least negative.
\\
\para{Correlation of popularity, difficulty and sentiment}
\Number{\plusplus{itemcounter}} 
There is a statically significant negative correlation between the popularity, difficulty, and negative sentiment of concurrency topics.
%
\\
\para{Race tool topics}
\Number{\plusplus{itemcounter}} Concurrency questions about data race tools can be grouped into 4 topics:
\topic{root cause identification}, \topic{general}, \topic{configuration \& execution}, and \topic{alternative use}. 
\Number{\plusplus{itemcounter}} Developers ask the most race tool questions 
about \topic{root cause identification} and the least about \topic{alternative use}.
\\
\para{Relation of our findings with previous work}  
\Number{\plusplus{itemcounter}}
Our findings relate to several findings of previous work, including the works by
Pinto \etal \cite{Pinto:15}, Barua \etal \cite{Barua:14}, Rosen and Shihab \cite{Rosen:16}, Lu \etal \cite{Lu:08}, 
Guzman \etal \cite{Guzman:14}, Sinha \etal \cite{Sinha:16}, Tourani \etal \cite{Tourani:14}, Wang \etal \cite{Wang:18}, and Bagherzadeh and Khatchadourian \cite{Bagherzadeh:19}.
While some of our findings agree with the findings of these previous works,
others sharply contrast.

\subsection{Extensions to Our Previous Work}
This work is a significantly revised and extended version of our previous work \cite{Ahmed:18}
with the following major extensions:
\begin{enumerate}[nosep]
  \item Addition of the new research question \bi{RQ5} and the sentiment analysis of 245,541 
  concurrency questions and answers for all 27 concurrency topics;
  \item Addition of the new research question \bi{RQ7} and the development of
  the race tool keywords, extraction and grouping of race tool questions
  into topics, and presenting a set of example questions
  that developers ask about race tool topics;
  \item Extension of our previous research question \bi{RQ6} with the analysis
  of 15 new correlations between sentiments, popularities, and difficulties of
  concurrency topics;
  \item Extension of our previous implications with the implications of our
  new findings for the practice, research, and education of concurrent software development;

  \item Extension of our previous relations with the relations of our new
  findings for
  \bi{RQ5}--\bi{RQ7} with the findings of the previous work; 
\item Extension of our previous data collection with the distribution of concurrency questions and answers over
different programming languages; 
  \item Addition of 10 more findings, to our 12 previous findings, and 10  more relations with the findings of 
  previous works, to our 16 previous relations;
  
  \item Extension of our related work with 
  previous works related to
  the sentiment analysis and usage of data race tools and techniques.
\end{enumerate} 

All the data used in this study are available at \url{https://goo.gl/uYCQPU}.

\subsection{Outline}
\secref{sec:methodology} presents our methodology for the collection and analysis of our data.
\secref{sec:results} discusses the answers to our research questions and
investigates the relation of our findings with the findings of the previous
work.
\secref{sec:implications} shows the implications of our findings for the practice, research, and education of concurrent software development.
\secref{sec:threats} discusses threats to the validity of our study and their mitigation strategies. 
\secref{sec:related} discusses related works. 
\secref{sec:conclusion} concludes the paper.

Throughout this paper, the difficulty of a topic is 
solely defined as the difficulty for finding  answers to the questions in that
topic.  



%% file: methodology-new.tex
\section{Methodology}
\label{sec:methodology}

\figref{fig-methodology} shows an overview of our methodology for the collection of concurrency questions and answers from \SO 
 and their analysis. 
\input{fig-methodology}


\subsection{Data Collection}
\newcounter{stepcounter}
\setcounter{stepcounter}{1}
\para{Step \arabic{stepcounter}: Download \SO dataset}
In this step of our methodology, we download the \SO dataset $\sopostset$ 
which is publicly available through Stack Exchange Data Dump 
\cite{Stackexchangedump}. 
The data set $\sopostset$ includes 
question and answer posts and their metadata.
The metadata of a post 
includes its identifier, its type (question or answer), title, body, tags, creation date, view count, score,
favorite count, and  the identifier of the accepted answer for the post, if the post is a question.
An answer to a question is accepted if the developer who posted the question 
marks it as accepted. The question can have at most 1 accepted answer. The
question can have 1 to 5 tags that are used to specify what the question is about.

Our dataset $\sopostset$ includes 38,485,046 questions and answers posted over a time span of over 
9 years from August 2008 to 
December 2017 by 3,589,412 developer participants of \SO. 
Among these posts, 14,995,834 (39\%) 
are questions and 23,489,212 (61\%) are answers. Among these answers
8,034,235 (21\%) are marked as accepted answers.

\stepcounter{stepcounter}
\para{Step \arabic{stepcounter}: Develop concurrency tag set}
In this step,
we develop a set of concurrency tags $\tagset$ to identify and extract concurrency
questions and answers in \SO. First, we start with a tag set
$\tagset_{\subscript{init}}$ that includes our initial concurrency tags. Second, we extract questions $\ipostset$
from our dataset $\sopostset$ whose tags match a tag in $\tagset_{\subscript{init}}$. Third, we
construct a set of candidate tags $\tagset$ by extracting tags of the questions
in $\ipostset$. Finally, we refine $\tagset$ by keeping tags that are significantly
relevant to concurrency and exclude others. We use two heuristics $\sig$
and $\rela$, that are also used by the previous work
\cite{Rosen:16,Yang:16,Bagherzadeh:19}, to measure the significance and
relevance of a tag $t$ in $\tagset$.
\begin{mathpar}
\text{(Significance)} \quad \sig = \dfrac{\text{number of questions with tag $t$ in $\ipostset$}}{\text{number of questions in $\ipostset$}}\\ 
\text{(Relevance)} ~~\quad \rela = \dfrac{\text{number of questions with tag $t$ in $\ipostset$}}{\text{number of questions with tag $t$ in $\sopostset$}} 
\end{mathpar}

A tag $t$ is significantly relevant to concurrency if its $\sig$ and $\rela$ are higher
than specific thresholds.
Our experiments using a broad range of thresholds for $\sig$ and $\rela$ show that 
$\mu=0.1$ and $\nu=0.01$ allow for a significantly relevant set of concurrency tags.
Our threshold values 
are in line with the values
used by previous work \cite{Yang:16,Barua:14,Bagherzadeh:19}.
\tabref{fig-tags} shows the concurrency tag set $\tagset$ for select significance and relevance 
values $\sig$ and $\rela$ with our tagset 
$\canset$ in gray.

We use the initial tag set $\tagset_{\subscript{init}}=\{\btag{multithreading}\}$ to construct the tag set  $\tagset$ below 
using our tag development approach.
To select the set of our initial tags in $\tagset_{\subscript{init}}$,
we manually inspect the top 100 most used tags on \SO and select all the tags that are related to concurrency.
Our set $\tagset_{\subscript{init}}$ includes a tag \btag{multithreading}, that
is used by  previous work \cite{Lu:08,Pinto:15} as well. 

\[
  \canset = \left\{ \begin{array}{p{10.5cm}}
\emph{concurrency locking multiprocessing multithreading mutex}
\emph{parallel-processing pthreads python-multithreading}
\emph{synchronization task-parallel-library thread-safety threadpool}        
  \end{array}\right\}
\]

Our concurrency tag set $\canset$ with 12 tags is broad and includes generic tags, such as \btag{concurrency} and \btag{synchronization}, 
and specific tags, such as 
\btag{thread safety} and \btag{thread pool}

\input{fig-tags}

\stepcounter{stepcounter}
\para{Step \arabic{stepcounter}: Extract concurrency posts}
In this step,
we extract \SO  questions whose tag set 
contains a tag in $\tagset$.
This set includes 156,777 questions and 249,662 answers, where 88,764 (36\%)
of these answers are accepted answers. 
To reduce noise, following previous work \cite{Barua:14,Rosen:16,Bagherzadeh:19}, we add questions and their accepted answers from this set to 
our set of concurrency post $\cpostset$ and discard unaccepted answers. 
The set of concurrency posts $\cpostset$ includes 156,777 questions and 88,764 accepted answers. In total,
$\cpostset$ includes 245,541 questions and answers. 

Using the tag set $\canset$ to extract concurrency questions
does not mean that concurrency questions cannot have other tags in addition to
the tags in $\canset$. 
In fact, our concurrency questions have 11,703 extra tags, such as 
\btag{asynchronous}, \btag{blocking}, and \btag{atomic}.

In
addition, our concurrency questions are related to a broad spectrum of
programming languages ranging from C languages, such as C, C++, C\#, and
Objective C, to Java and JavaScript and from Visual Basic to Go and
Haskell.
\figref{fig-languages} shows the number of our concurrency questions for
different programming languages. To extract these languages, we manually inspect
top 200 most used tags in $\cpostset$ and select tags that identify a
programming language.

\input{fig-languages}

\stepcounter{stepcounter}
\para{Step \arabic{stepcounter}: Develop race tool keyword set}
In this step, we develop a set of keywords $\kset$ to identify and extract concurrency questions
that are about tools and techniques that are used to find and
fix data race bugs.
%
%
We develop a keyword set instead of a tag set 
because of the following reasons. First, there are no \SO tags that
are specifically designed to represent data race tools and techniques.
Second, most \SO questions with the
generic data race tags \cite{DataraceTags}, such as \btag{data-race} and
\btag{race-condition}, are not about the usage of data race tools and
techniques.

To construct the keyword set $\kset$, we include the names of both industrial 
as well as academic data race tools and techniques that are discussed in 
the previous work. 
%
For industrial tools, we include Google's \btag{ThreadSanitizer} (\btag{TSan})
and \btag{ThreadSafety}  (\btag{annotalysis}) \cite{Sadowski:14}, Intel's \btag{Inspector} \cite{IntelInspector}, 
and Valgrind's \btag{Helgrind} and \btag{DRD} \cite{Helgrind,DRD}. 
For academic tools and techniques, we add 29 tools and techniques from a recent and comprehensive survey on  data race tools and techniques \cite{Hong:15}.
We exclude tools and techniques with no specific names or overly generic names such as \btag{HAVE}.  
%
\[
  \kset = \left\{ \begin{array}{p{10.5cm}}
        \emph{ThreadSanitizer TSan ThreadSafety annotalysis Helgrind}
		\emph{DRD 'Intel Inspector' Acculock Eraser FastTrack LiteRace Racer}
        \emph{RaceTrack RACEZ SOS TRaDe Atom-Aid AtomRace AVIO}
        \emph{CTrigger DefUse FALCON McPatom PENELOPE MultiRace}
        \emph{MUVI-Eraser AtomTracker ColorSafe COPPER Marathon}
        \emph{MUVI-AVIO Veldrome Chord RacerX RccJava RELAY}
  \end{array}\right\}
\]
The keyword set $\kset$ with 36 keywords covers a broad set of static and dynamic 
tools and techniques for finding and fixing data race bugs. 
To illustrate, \btag{Chord} and \btag{RacerX} are static whereas \btag{ThreadSanitizer} and \btag{ThreadSafety} are dynamic.
A static race detection technique analyzes a program without running it, whereas a dynamic technique 
analyzes the runtime information of the program.

\stepcounter{stepcounter}
\para{Step \arabic{stepcounter}: Extract race tool posts}
In this step, we first extract concurrency questions from $\cpostset$ that their title or body include a keyword in $\kset$. 
This set includes 446 questions. Second, we manually study these questions to exclude the questions that are not about 
data race tools. 
This is because a keyword in $\kset$ may refer to different tools and techniques 
used for 
different purposes.
To illustrate, \btag{FALCON} is not only a data race tool but also a Python web development framework.
Similarly, \btag{Helgrind} can not only be used for race detection but also deadlock.
Our final set $\uset$ of race tool posts includes 82 questions. Interestingly, the questions in $\uset$ are related to the 
following 6 keywords in $\kset$ and 
there are no questions in  $\uset$ for the rest of keywords.
\[
  \left\{ \begin{array}{p{11cm}}
\emph{ThreadSanitizer TSan Helgrind DRD 'Intel Inspector' CTrigger RacerX}  
\end{array}\right\}
\]


%
   

%

\stepcounter{stepcounter}
\para{Step \arabic{stepcounter}: Preprocess concurrency posts}
In this step, we preprocess the set of our concurrency posts $\cpostset$ to
reduce the noise for sentiment analysis and topic modeling in the next steps.
%
For sentiment analysis, we preprocess $\cpostset$ 
by removing code snippets, 
enclosed in $\left<code\right>$$\left<\backslash code\right>$ tags, HTML tags, such as $\left<p\right>$ $\left</p\right>$,
numbers, punctuation marks, non-alphabetical characters, and URLs.
For topic modeling, we further preprocess $\cpostset$ 
by removing
stop words, such as ``a'', ``the'', and ``is'', and reduce words to their base representations. For example,
  ``reading'', ``read'' and ``reads'' all reduce to their base ``read''. Reducing words to their bases allows grouping of words with 
  similar meanings together. 
For stop words 
and word reduction, we use  \Mallet's popular list of stop words \cite{Porter:97} and Porter stemming algorithm \cite{Porter:97}, respectively.
We do not 
reduce words for sentiment analysis because an unreduced  
word may convey important
sentiment information \cite{Mantyla:17,Souza:17,Sinha:16}.

\subsubsection{Data collection for concurrency questions and answers}
Our data collection steps are in line with the best practices of
previous work.
Previous work uses tags, keywords, and significance and relevance heuristics
often to identify security \cite{Yang:16,Meng:18}, chatbot \cite{Abdellatif:20},
mobile \cite{Rosen:16}, big data \cite{Bagherzadeh:19}, and deep learning
\cite{Islam:19,Islam:20} questions and answers from \SO.

\subsection{Data Analysis}

\stepcounter{stepcounter}
\para{Step \arabic{stepcounter}: Model and label concurrency topics}
In this step, 
%
we use  \Mallet \cite{McCallum:02} with latent Dirichlet allocation (LDA)  topic modeling \cite{Blei:03}
to group our concurrency posts  $\cpostset$ into topics. 
%
In our topic model, a concurrency post is a probabilistic distribution of several topics
with different proportions. 
A post has a dominant topic that covers the biggest
proportion of the text of the post. A topic is a set of frequently co-occurring
words that approximates a real-world concept. 
To produce the model, we treat each individual question and accepted answer  as an
individual document. \Mallet processes 245,541 questions and answers documents.

\Mallet groups
posts into $K$ topics after $I$ grouping iterations and returns a set of
topics and their proportions for each post along with a set of words for
each topic. 
Our experiments with a broad range of values 
show that $K=30$ and $I=1,000$  
allow for sufficiently granular topics. 
These values are in line with values used by previous work
\cite{Bajaj:14,Rosen:16,Yang:16,Bagherzadeh:19}.
MALLET uses hyperparameters $\alpha$ and $\beta$ to control the distribution of words
in topics and distributions of topics in posts. 
Following previous work \cite{Bajaj:14,Rosen:16,Yang:16,Bagherzadeh:19}, we use standard values
50/K and 0.01 for these hyperparameters. We use this standard values 
because as Bigger et al.
\cite{Biggers:14} show "$\alpha$ and $\beta$ have little influence on the accuracy of the LDA [topic modeling]".

\Mallet groups concurrency posts  $\cpostset$ into topics $t$ that are a set
of words. However, \Mallet cannot decide about the meaning of the topics and
label them with names useful for human beings.
To illustrate, the word set $w = \{task,$ $execute,$ $async,$ $complete,$ $run,$ $cancel,$ $wait,$ $asynchronous,$ $parallel,$ $schedule\}$ 
represents a topic in \Mallet.
Following previous work \cite{Rosen:16,Yang:16,Barua:14,Bajaj:14,Nadi:16,Bagherzadeh:19}, we 
use an open card sort technique \cite{Fincher:05} to label the topics. 
%
%
In an open card sort, there are no predefined topics and participants develop their own topics during the sorting and labeling process.
To label a topic, we find it sufficient to manually inspect the top 20 words in the set of words for
the topic and read through 15 random posts for the topic to decide for
a label that best explains the words and posts of the topic. 
To illustrate, we label the word set $w$ with \topic{task parallelism}.
The numbers of topic words and random posts used to label
a topic are in line with previous work
\cite{Bagherzadeh:19,Bajaj:14,Barua:14,Pinto:14,Pinto:15,Rosen:16,Yang:16}.
\tabref{fig-topics} shows our 27 concurrency topics, their names,
and top 10 reduced words. 

During the labeling process, we merge topics 1 and 15 into \topic{basic concepts} topic due to 
close similarities between their topic words and questions.
Similarly, we merge topics 5 and 28 into \topic{object-oriented concurrency}. 
%
We remove topic 12  
that is about synchronization between local and remote repositories in version control systems
such as Git and is not about concurrency. 

%

During the card sorting process for concurrency topics, the first and second authors individually assign labels to topics and reiterate and refine topics 
until they agree on topic labels.
The first author is a Programming Languages and Software Engineering professor 
with extensive expertise in concurrent and event-based systems
\cite{Bagherzadeh:20,Khatchadourian:20,Khatchadourian:18,Bagherzadeh:19,Ahmed:18,Bagherzadeh:17,Long:16,Bagherzadeh:15}
and the second author is a graduate student with coursework in concurrent and distributed systems.
The first author also has several years of industrial experience  as a Software Engineer.


\stepcounter{stepcounter}
\para{Step \arabic{stepcounter}: Construct concurrency topic hierarchy}
In this step, we use a card sorting process, similar to the step 7 and with the
same authors, to
construct the topic hierarchy by repeated grouping of similar topics into categories, 
and lower-level categories into higher-level categories.
There are no predefined 
categories and and
participants develop their own categories during the sorting and labeling
process. \tabref{fig-topics} and  \figref{fig-posts-sunburst} show the textual and pictorial representations of the topic hierarchy.  
To illustrate, \topic{thread life cycle management} and \topic{thread
scheduling} topics are grouped into the lower-level  category \topic{multithreading}, where \topic{multithreading} itself is grouped into the higher-level category \topic{concurrency models}.
The category \topic{concurrency models} includes other lower-level categories such as 
\topic{multiprocessing} and \topic{parallel computing}.
  
\stepcounter{stepcounter}
\para{Step \arabic{stepcounter}: Determine concurrency topic popularity}
In this step, we use 3 well-known metrics that are used often by the previous
work to measure the popularity of a concurrency topic.
%
The first metric, that is used by the previous work
\cite{Bajaj:14,Rosen:16,Yang:16,Nadi:16,Bagherzadeh:19,Abdellatif:20}, 
is the average number
of views for questions with the topic as their dominant topic. 
This metric includes views by both registered users and visitors of \SO. 
The inclusion of views by visitors is important because in \SO there are
many more visitors than there are registered users \cite{Mamykina:11}.
The second metric, that is used by the previous work
\cite{Yang:16,Pinto:15,Nadi:16,Bajaj:14,Bagherzadeh:19,Abdellatif:20} 
is the average number of questions of the topic marked as favorite by users.
The third metric, that is used by the previous work
\cite{Yang:16,Pinto:15,Nadi:16,Bajaj:14,Bagherzadeh:19,Abdellatif:20}, is the
average score of questions of the topic.
Intuitively, a topic with a higher number of views and favorites and a higher score is more popular.  
\tabref{fig-topic-pop-table} shows the popularity measurements of concurrency topics.

\stepcounter{stepcounter}
\para{Step \arabic{stepcounter}: Determine concurrency topic difficulty}
%
We use 2 well-known metrics that are used often by previous work to measure the
difficulty of a concurrency topic.
The first metric, that is used by the previous work
\cite{Yang:16,Rosen:16,Treude:11,Bagherzadeh:19,Abdellatif:20},  is the
percentage of questions of the topic that have no accepted answers .
And the second metric, that is used by the previous work
\cite{Yang:16,Rosen:16,Bagherzadeh:19,Abdellatif:20}, is the average median time
needed for a question to receive an accepted answer.
Intuitively, a topic with  higher percentage of questions with no accepted
answers and taking longer to receive accepted answers is more difficult.
\tabref{fig-topic-diff-table} shows the difficulty measurements of our concurrency topics. 

\stepcounter{stepcounter}
\para{Step \arabic{stepcounter}: Determine concurrency topic sentiment}
We use \SentiSD \cite{Calefato:18} to determine 
the sentiment of our concurrency topics. 
\SentiSD 
uses a combination of 
lexical and semantic features to determine the sentiment polarity of a concurrency post and 
classify it as positive, negative, or neutral.
For example, \SentiSD classifies the sentence 
``I am so used to thinking about solutions in a linear/serial/OOP/functional way
that I am struggling to cast any of my domain problems in a way that merits
using concurrency'' from a \SO post as negative whereas it classifies the
sentence ``what challenges promote the use of parallel/concurrent architectures?'' as
neutral. 
 \SentiSD is
specifically trained 
to analyze communications of software developers. Previous work
\cite{Novielli:18} shows that \SentiSD provides the best
performance, in comparison to other sentiment analysis tools such SentiStrength \cite{Thelwall:12}, SentiCR \cite{Ahmed:17}, and
SentiStrengthSE \cite{Islam:17},
when analyzing \SO questions and answers.
The sentiment of a concurrency topic is the percentages of the positive, negative, and neutral polarities of its questions and answers.
\tabref{fig-topic-sent-table} shows the sentiment measurements of concurrency
topics. 


\stepcounter{stepcounter}
\para{Step \arabic{stepcounter}: Determine popularity, difficulty, and sentiment correlations}
In this step, 
we use Kendall correlation to identify if there are any correlations between 
popularity, difficulty, and sentiment of our concurrency topics.
%
We use Kendall because it is less sensitive to outliers and is relatively more stable. 
In total, we investigate 21 correlations between 3 popularity, 2 difficulty, and
3 sentiment metrics of our concurrency topics.


\stepcounter{stepcounter}
\para{Step \arabic{stepcounter}: Label race tool topics}
In this step, we use the card sorting technique, similar to steps 7 and 8, 
to group race tool questions in $\uset$ into race tool topics.
During the card sorting process, the first and 
fourth authors individually assign labels to topics and reiterate and refine the 
topics until they agree on topic labels.
The fourth author is a Software Engineering professor 
with extensive expertise in parallel systems \cite{Khatchadourian:20,Khatchadourian:19}. 
The fourth author also has several years
of industrial experience as a Software Engineer.

\stepcounter{stepcounter}
\para{Step \arabic{stepcounter}: Investigate the relation of our findings
with the findings of previous work} Throughout steps  7--13 of our analysis, we
investigate the relation of our findings 
in each of these steps with the findings of previous work. 

\subsubsection{Data analysis of concurrency questions and answers}
Our data analysis steps are in line with the best practices of
previous work.
Previous work uses card sorting and manual analysis often to
group security \cite{Yang:16,Meng:18}, chatbot \cite{Abdellatif:20},
mobile \cite{Rosen:16}, big data \cite{Bagherzadeh:19}, and deep learning
\cite{Islam:19,Islam:20} questions and answers from \SO.
In addition, previous work uses the average number of views, favorites, and scores as popularity metrics and 
the percentage of questions with no answers and median time to receive  
accepted answers as difficulty metrics to 
measure the popularity and difficulty of security \cite{Yang:16,Meng:18}, chatbot \cite{Abdellatif:20},
mobile \cite{Rosen:16}, and big data \cite{Bagherzadeh:19} topics.

%

%
%



%% file: fig-methodology.tex
\begin{figure}[h!]
\captionskipneg
\centering
\includegraphics[scale=.5]{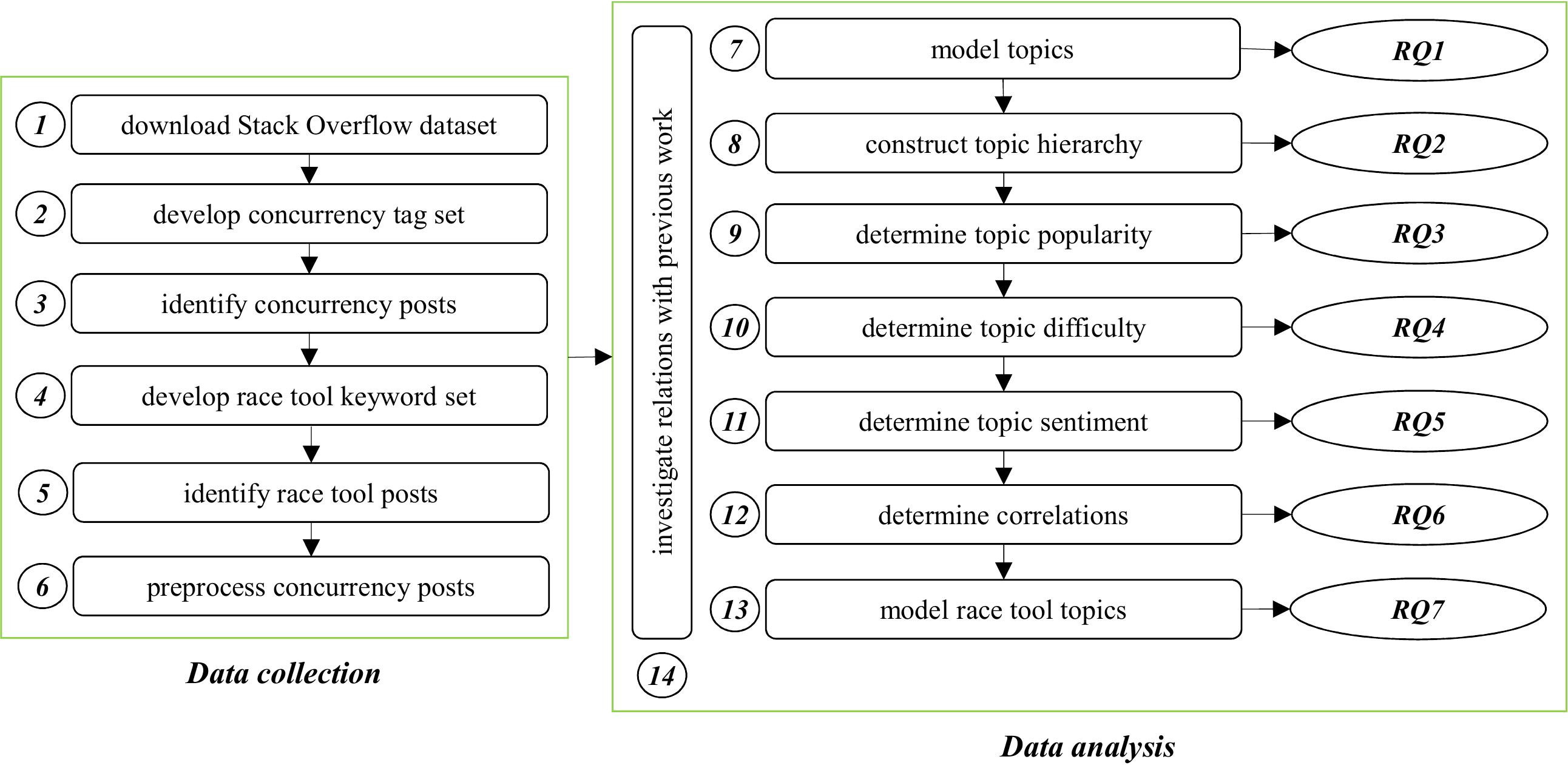}
\negspace
\caption{
An overview of the methodology of our study.
%
}
\label{fig-methodology}
\end{figure}

%% file: fig-tags.tex
\begin{table}[h!]
\captionskiptableneg
\caption{Concurrency tags for select relevance and significance threshold values. Our concurrency tag set $\canset$ is in gray.}
\label{fig-tags}
\negspace
\begin{small}
\newcolumntype{P}[1]{>{\centering\arraybackslash}p{#1}}

\setlength{\tabcolsep}{1mm}

\begin{tabular*}{\textwidth}{@{}l p{9.3cm} c}
\rowcolor[HTML]{333333} {\color[HTML]{FFFFFF} \textit{\textbf{$(\mu,\nu)$}}} & \rule{0pt}{5pt} {\color[HTML]{FFFFFF} \textit{\textbf{Tag set $\tagset$}}}                                                                                                                                                                                                                                                                                 & {\color[HTML]{FFFFFF} \textit{\textbf{No.}}} \\
(0.3, 0.015)                                        & concurrency \msrspace multithreading \msrspace pthreads \msrspace thread-safety \msrspace threadpool                                                                                                                                                                                                                                                                 & 5                                                    \\
(0.3, 0.01)                                         & concurrency \msrspace multithreading \msrspace mutex \msrspace pthreads \msrspace python-multithreading \msrspace thread-safety \msrspace threadpool                                                                                                                                                                                                                                   & 7                                                     \\
(0.3, 0.005)                                        & backgroundworker \msrspace concurrency \msrspace executorservice \msrspace multithreading \msrspace mutex \msrspace pthreads \msrspace python-multithreading \msrspace runnable \msrspace semaphore \msrspace synchronized \msrspace thread-safety \msrspace threadpool                                                                                                                                                             & 12                                                    \\
\hline
(0.2, 0.015)                                        & concurrency \msrspace locking \msrspace multithreading \msrspace pthreads \msrspace synchronization \msrspace thread-safety \msrspace threadpool                                                                                                                                                                                                                                       & 7                                                     \\
(0.2, 0.01)                                         & concurrency \msrspace locking \msrspace multithreading \msrspace mutex \msrspace pthreads \msrspace python-multithreading \msrspace synchronization \msrspace task-parallel-library \msrspace thread-safety \msrspace threadpool                                                                                                                                                                                  & 10                                                    \\
(0.2, 0.005)                                        & atomic \msrspace backgroundworker \msrspace concurrency \msrspace deadlock \msrspace executorservice \msrspace grand-central-dispatch \msrspace locking \msrspace multithreading \msrspace mutex \msrspace openmp \msrspace pthreads \msrspace python-multithreading \msrspace runnable \msrspace semaphore \msrspace synchronization \msrspace synchronized \msrspace task-parallel-library \msrspace thread-safety \msrspace threadpool \msrspace wait                                                    & 20                                                    \\
\hline
(0.1,0.015)                                        & concurrency \msrspace locking \msrspace multithreading \msrspace parallel-processing \msrspace pthreads \msrspace synchronization \msrspace thread-safety \msrspace threadpool                                                                                                                                                                                                                  & 8                                                     \\
\rowcolor[HTML]{EFEFEF} {\color[HTML]{333333} (0.1, 0.01)}                  & \rule{0pt}{5pt}{\color[HTML]{333333} concurrency \msrspace locking \msrspace multiprocessing \msrspace multithreading \msrspace mutex \msrspace parallel-processing \msrspace pthreads \msrspace python-multithreading \msrspace synchronization \msrspace task-parallel-library \msrspace thread-safety \msrspace threadpool}                                                        & 12                                                    \\
(0.1, 0.005)                                        & atomic \msrspace backgroundworker \msrspace concurrency \msrspace deadlock \msrspace executorservice \msrspace grand-central-dispatch \msrspace locking \msrspace multiprocessing \msrspace multithreading \msrspace mutex \msrspace openmp \msrspace parallel-processing \msrspace pthreads \msrspace python-multithreading \msrspace queue \msrspace runnable \msrspace semaphore \msrspace synchronization \msrspace synchronized \msrspace task \msrspace task-parallel-library \msrspace thread-safety \msrspace threadpool \msrspace wait & 24                                                    \\
\hline
(0.05, 0.015)                                       & asynchronous \msrspace c++11 \msrspace concurrency \msrspace locking \msrspace multithreading \msrspace parallel-processing \msrspace pthreads \msrspace sockets \msrspace synchronization \msrspace thread-safety \msrspace threadpool                                                                                                                                                                                    & 11                                                    \\
(0.05, 0.01)                                        & android-asynctask \msrspace asynchronous \msrspace boost \msrspace c++11 \msrspace concurrency \msrspace locking \msrspace multiprocessing \msrspace multithreading \msrspace mutex \msrspace parallel-processing \msrspace pthreads \msrspace python-multithreading \msrspace sockets \msrspace synchronization \msrspace task-parallel-library \msrspace thread-safety \msrspace threadpool \msrspace timer                                                                             & 18                                                    \\
(0.05, 0.005)                                       & android-asynctask \msrspace async-await \msrspace asynchronous \msrspace atomic \msrspace backgroundworker \msrspace boost \msrspace c++11 \msrspace concurrency \msrspace deadlock \msrspace executorservice \msrspace grand-central-dispatch \msrspace locking \msrspace multiprocessing \msrspace multithreading \msrspace mutex \msrspace openmp \msrspace parallel-processing \msrspace process \msrspace pthreads \msrspace python-multithreading \msrspace queue \msrspace runnable \msrspace semaphore \msrspace sockets \msrspace synchronization \msrspace synchronized \msrspace task \msrspace task-parallel-library \msrspace thread-safety \msrspace threadpool \msrspace timer \msrspace wait & 32\\
\end{tabular*}
\end{small}
\end{table}

%% file: fig-languages.tex
\begin{figure}[H]
\captionskipneg
\centering
\includegraphics[scale=.75]{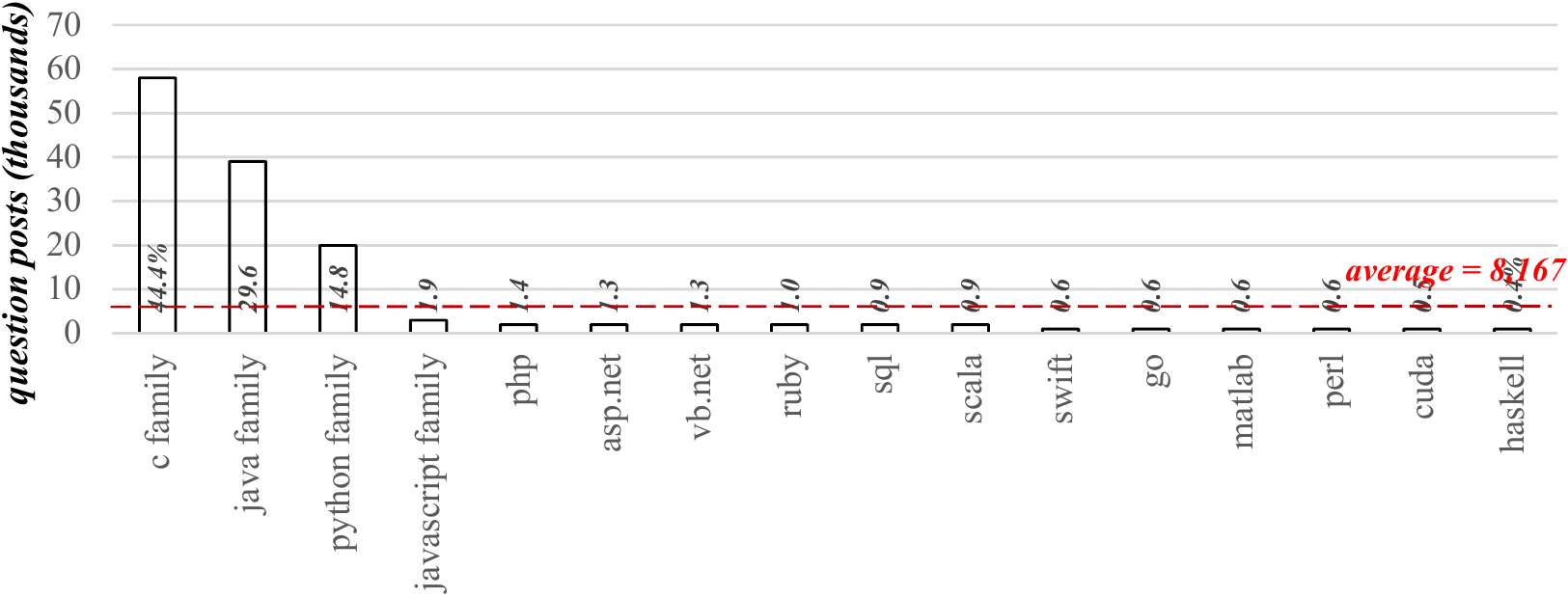}
\negspace
\caption{Programming languges of our concurrency questions and their numbers.}
\label{fig-languages}
\end{figure}

%% file: results.tex
\newcounter{findingcounter}
\setcounter{findingcounter}{0}

\section{Results}
\label{sec:results}

In this section, we present and discuss the results of our study for 
research questions 
\bi{RQ1}-\bi{RQ7}, 
investigate the relation of our findings
with the findings of the previous work, and 
present a set of
example questions that developers
ask for each of our concurrency and race tool topics and categories.

\subsection{RQ1: Concurrency Topics}

\input{fig-topics}

\tabref{fig-topics} shows the concurrency topics that developers ask questions
about on \SO. These topics are determined in step 7 of our analysis.
%
According to \tabref{fig-topics},
developers ask questions about a broad spectrum of concurrency topics.
These questions can be categorized into 27 concurrency topics 
ranging 
from \topic{thread pool} to \topic{parallel computing}, \topic{mobile concurrency} to \topic{web concurrency}, and
\topic{memory consistency} to \topic{runtime speedup}.

The meaning of a concurrency topic is best understood by looking at 
the questions that belong to the topic. 
To illustrate, the following is a question in the \topic{thread pool} topic.
In this question,
%
the developer is asking 
how to implement a thread pool  
where the size of the pool can change based on the number of its jobs.
The \SO identifier for this question is \emph{11249342} and 
it 
can be accessed at \url{https://stackoverflow.com/questions/11249342}.
This question has been viewed 13,448 times, more than 6 times
the average number of views for a \SO question. The
average number of views for a \SO question is 2,154 \cite{Bagherzadeh:19}.
%
A thread is an abstraction for the execution of a computation that allows to control its
concurrency 
by running, suspending, and resuming the execution of the computation. 
A thread pool is an abstraction for a set of threads that allows for their creation, allocation, and deallocation.     
\bigquoted{11249342}{Creating a dynamic (growing/shrinking) thread pool}
{I need to implement a thread pool in Java (java.util.concurrent) whose number of threads is at some minimum value when idle, grows up to an upper bound (but never further) when jobs are submitted into it faster than they finish executing, and shrinks back to the lower bound when all jobs are done \ldots
How would you implement something like that? \ldots}


%

Similarly, the following is a question in the \topic{web concurrency} topic.
In this question,
%
the developer is asking
how to send emails using background threads in their web application where the background thread 
prevents blocking of the main thread and therefore does not force the user to wait until the email is sent. 
Classic ASP (Active Server Pages) 
is a scripting language to write server-side web applications. 
Ajax is a client-side web development technique 
for asynchronous communication between a web client and server. 
In an asynchronous request, the client does not block to wait for the response from the server and can immediately continue 
its computation after the request is made.        
\bigquoted{17052243}{How to perform multithreading/background process in classic ASP}
{I need to send emails via a background job on a classic-ASP app so the user doesn't have to wait for a slow webserver to complete sending the email.
I know I can use Ajax to generate two separate requests, but I'd rather not require Javascript. Plus, I suspect there's a better way to pull this off. Ideas?
}


Finally, the following is a question in the \topic{basic concepts} topic. In
this question, the developer asks about the basic motivations behind the
need for concurrency and concurrency architectures. 
\bigquoted{541344}{What challenges promote the use of parallel/concurrent architectures?}
{.. I am so used to thinking about solutions in a linear/serial/OOP [Object-oriented Programming]/functional way that I am struggling to cast any of my domain problems in a way that merits using concurrency.}


\finding{\plusplus{findingcounter}}
{Questions that concurrency developers ask can be grouped into 27 concurrency
topics
ranging 
from \topic{thread pool} to \topic{parallel computing}, \topic{mobile concurrency} to \topic{web concurrency}, and
\topic{memory consistency} to \topic{runtime speedup}.
}

\input{fig-topic-question}

\figref{fig-topic-question} shows the number of questions that developers ask about our concurrency topics. 
%
According to \figref{fig-topic-question}, 
the numbers of questions that developers ask for concurrency topics are not uniform.
Developers ask the most number of questions (\percent{8}) about \topic{basic concepts} and  
the least number of questions (\percent{1}) about \topic{event-based concurrency}.


\coincidence{(Pinto \etal and Rosen and Shihab)}
Pinto \etal \cite{Pinto:15} study the 250 most popular concurrency questions on
\SO and group them into several themes. 
Our observation that the concurrency topic \topic{basic concepts} has the most number of questions 
is similar to Pinto \etal's observation that ``most of them [concurrency
questions] are related to basic concepts''.
Our observation is also in line with the general understanding that the concurrency
and even its basics remains challenging for developers.
Bagherzadeh and Khatchadourian \cite{Bagherzadeh:19} study and categorize \SO questions and answers related to
big data software development into several big data topics and categories including the \topic{basic concepts} topic. 
Our \topic{basic concepts} topic overlaps with Bagherzadeh and Khatchadourian's
\topic{basic concepts}.
Interestingly, both concurrency and big data developers ask more than average number of questions about \topic{basic concepts}.


\finding{\plusplus{findingcounter}}
{
Developers ask the most 
questions about \topic{basic concepts} inline with the general
understanding that the concurrency and even its basics remains challenging for
developers. Developers ask 
the least 
about \topic{event-based concurrency}.
}

\subsection{RQ2: Concurrency Topics Hierarchy}
\figref{fig-posts-sunburst} shows the hierarchy of our concurrency topics.
This hierarchy is constructed in step 8 of our study.
In this figure, concurrency topics are in gray and their categories are in white with the percentages of their questions. The
hierarchy expands outwards from higher-level to lower-level categories to topics.


%

\input{fig-posts-sunburst}

According to \figref{fig-posts-sunburst}, 
questions that concurrency developers ask can be grouped into a hierarchy with 
8 high-level categories: 
\topic{concurrency models}, \topic{programming paradigms}, \topic{correctness},
\topic{debugging}, \topic{basic concepts}, \topic{persistence}, \topic{performance}, and \topic{graphical user interface (GUI)}.
In addition, the number of questions that developers ask in each category is not
uniform.
Developers ask the most \ppercent{28} about the \topic{concurrency models} and the least \ppercent{5} about \topic{GUI}.  
Interestingly, developers ask more \ppercent{12} about \topic{correctness} of
their concurrent programs than \topic{performance} \ppercent{7}.
This is in line with the general tradeoff between the performance advantages of
concurrency and its correctness issues \cite{Bagherzadeh:17,Bagherzadeh:15,Long:16,Lee:06}.

\finding{\plusplus{findingcounter}}
{Questions that 
developers ask can be grouped into a hierarchy with 
8 high-level categories: 
\topic{concurrency models}, \topic{programming paradigms}, \topic{correctness}
\topic{debugging}, \topic{basic concepts}, \topic{persistence}, \topic{performance}, and \topic{GUI}.
}

\finding{\plusplus{findingcounter}}
{
Developers ask the most questions 
about \topic{concurrency models} category and the least 
about \topic{GUI}.   
}

\finding{\plusplus{findingcounter}}
{
Developers ask more 
about \topic{correctness} than \topic{performance}, 
in line with the correctness and performance tradeoffs of concurrent software.
%
}

In the following, we discuss each higher-level concurrency category and
its lower-level categories and topics. In addition, we present a set of
example questions for each concurrency topic and category.


\subsubsection{Concurrency Models}
Concurrency models are  
concerned about concurrency abstractions (e.g. threads and processes)
and execution models (e.g. multicore and single core executions).
The \topic{concurrency models} category includes five lower-level categories among which 
 \topic{multithreading} alone contains more than half of the questions that developers ask in this category.
This is in line with the general understanding that
multithreading is the defacto dominant concurrency model.  
Developers ask the most (\percent{54}) about \topic{multithreading} and the least (\percent{6}) about 
\topic{producer-consumer concurrency}, when asking questions about concurrency
models.

In \topic{multithreading}, 
developers ask questions with titles like 
\vquestion{thread life cycle management}{How to safely destruct Posix thread pool in C++?} and
\vquestion{thread pool}{Is there a way to create a pool of pools using the Python workerpool module?}. 
The topic of a question is included inside parentheses.
In
\topic{producer-consumer concurrency}, developers ask questions with titles like
\squestion{producer-consumer concurrency}{producer Consumer with BlockingQueues in Java EE as background task}.
%

\coincidence{}
%
Our \topic{multithreading} category and its \topic{thread life cycle management} topic 
overlap with Pinto \etal's \cite{Pinto:15} \topic{threading} and \topic{thread
life cycle} themes.
Rosen and Shihab \cite{Rosen:16} study and categorize \SO questions related to mobile development 
into several mobile topics including \emph{threading}. 
Our \topic{multithreading} category  overlaps with their mobile \emph{threading}
topic.

\finding{\plusplus{findingcounter}}
{
Developers ask the most 
about \topic{multithreading}, when asking about \topic{concurrency models}, 
in line with the general understanding that multithreading is the defacto
dominant concurrency model.
 Developers ask the least 
about \topic{producer-consumer concurrency}.
}

\subsubsection{Programming Paradigms}
Programming paradigms are  concerned about programming abstractions (e.g. objects and events), platforms (e.g. web and mobile), and 
patterns (e.g. producer and consumer).
The \topic{programming paradigms} category includes five lower-level categories among which 
\topic{object-oriented concurrency} alone contains more than a third of questions that developers ask in this category.
This is in line with the general understanding that object-orientation is a
dominant programming paradigm.
Developers ask the most (\percent{35}) about 
\topic{object-oriented concurrency} and the least (\percent{6}) about
\topic{event-based concurrency}, when asking about programming paradigms.

In the \topic{object-oriented concurrency} and \topic{event-based concurrency} topics, developers ask
questions with titles like \vquestion{object-oriented concurrency}{Is it better to synchronize object from inside of the class that encapsulates access it or from outside?} 
and \vquestion{event-based concurrency}{What type of timer event should I use for a background process when my timer fires very quickly?}.

\coincidence{}
Barua \etal \cite{Barua:14} study and categorize all \SO questions and answers into several general topics including
web and mobile development. Our concurrency topics \topic{mobile concurrency}
and \topic{web concurrency} overlap with their general  web development and
mobile development topics.
Our \topic{mobile concurrency} topic overlaps with Pinto \etal's
\cite{Pinto:15} observation that ``concurrent programming has reached mobile developers''. 
In 
Pinto \etal's  \cite{Pinto:15} study, \percent{9} of \SO questions are related to mobile development.
In contrast, according to \figref{fig-topic-question}, 
only \percent{4.6} of our concurrency questions are in \topic{mobile concurrency}
topic.   
  
\finding{\plusplus{findingcounter}}
{
Developers ask the most 
about 
\topic{object-oriented concurrency}, when asking about \topic{programming
paradigms}, in line with the 
general understanding that object-orientation is a dominant concurrent
programming paradigm.
Developers ask 
the least 
%
about \topic{event-based concurrency}.
}

\subsubsection{Correctness}
Correctness is concerned with the prevention of data corruption for concurrently accessed (e.g. read and write) data using 
mechanisms like locking, consistent memory models, and thread-safe data structures and programming patterns.
\topic{Correctness} questions are divided almost evenly among its 
\topic{thread safety}, \topic{locking}, \topic{concurrent collections}, and \topic{memory consistency} topics.

In \topic{correctness} category, developers ask questions like
\vquestion{thread safety}{How to make factory [pattern] thread safe?}, 
\vquestion{locking}{Is there a way to lock 2 or more locks or monitors atomically?},
\vquestion{concurrent collections}{Threadsafe dictionary that does lookups with minimal locking},   
and \vquestion{memory consistency}{Atomic read-modify-write in C\#}.
 
\coincidence{}
Lu \etal \cite{Lu:08} categorize concurrency bug patterns and their fixes.
Our \topic{memory consistency} topic overlaps with their observations that most
concurrency bugs are atomicity violation bugs where the
"desired serializability among multiple memory accesses is violated" and order violation bugs where 
the "desired order between two (groups of) memory accesses is flipped."
Our \topic{locking} topic overlaps with their designation that locking is one of
the main fixes for concurrency bugs to ensure correctness. 
In addition, our \topic{correctness} category and its \topic{locking} topic
overlap with Pinto \etal's \cite{Pinto:15} \topic{correctness} and
\topic{locking} themes.
Similarly, our \topic{concurrent collections} topic overlaps with Pinto \etal's
\topic{concurrent libraries} theme.

\finding{\plusplus{findingcounter}}
{
%
\topic{Correctness} questions are divided almost evenly between
\topic{thread safety}, 
\topic{locking}, 
\topic{concurrent collections}, 
and \topic{memory consistency} topics.

}

\subsubsection{Basic Concepts}
\topic{Basic concepts} are about  
both theoretical and practical aspect of concurrency and includes questions with 
titles like \squestion{}{How many threads are involved in deadlock?},
\squestion{}{What is a race condition?}, 
\squestion{}{Lock, mutex, semaphore... what's the difference?}, 
and \squestion{}{Java: notify() vs. notifyAll() all over again}.
  
\coincidence{}  
Our \topic{basic concepts} category overlaps with Pinto \etal's \cite{Pinto:15}
\otopic{theoretical concepts} and \otopic{practical concepts} themes.

\subsubsection{Debugging}
Debugging is concerned about finding and fixing concurrency bugs which manifest either in the behavior or output of programs.
Debugging questions are almost evenly divided among its topics \topic{irreproducible behavior}
and \topic{unexpected output}, which includes
question with titles like 
\squestion{irreproducible behavior}{Trace non-reproducible bug in C++} 
and \squestion{unexpected output}{Synchronized codes with unexpected outputs}.

\coincidence{}
Our \topic{debugging} category overlaps with Bagherzadeh and Khatchadourian's
\cite{Bagherzadeh:19} big data \topic{debugging} topic.
Our \topic{irreproducible behavior} topic in \topic{debugging} category overlaps
with Lu \etal's  \cite{Lu:08} observation that some ''concurrency bugs are very difficult to repeat''.

\finding{\plusplus{findingcounter}}
{\topic{Debugging} questions are divided almost evenly
between \topic{irreproducible behavior} and \topic{unexpected output} topics.}

\subsubsection{Persistence} 
Persistence is about storing and retrieving of data using persistence management systems (e.g. databases management systems, 
entity/object persistence\footnote{An entity management system automates the serialization of objects (entities) for storage in a database.}
or file systems).  
\topic{Persistence} includes three topics among which \topic{database management systems}  includes near half of persistence questions.
The \topic{persistence} category includes question 
with titles 
like
\vquestion{database management system}{How do I lock read/write to MySQL tables so that I can select and then insert without other programs
reading/writing to the database?},
\vquestion{file management}{Is this is correct use of mutex to avoid concurrent modification to file?},
and \vquestion{entity management}{Save entity using threads with JPA [Java Persistence API] (synchronized)}.
 
\coincidence{} 
Our \topic{database management systems} topic overlaps with Barua \etal's
\cite{Barua:14} general \otopic{MySQL} topic.

\finding{\plusplus{findingcounter}}
{Nearly half of \topic{persistence} questions are about \topic{database management systems}.}

\subsubsection{Performance}
Performance is about speeding up execution of programs (e.g. data scraping programs\footnote{A data scraping program downloads data from remote web URLs and stores it locally.}).
\topic{Performance} includes two topics. These topics includes questions with titles like 
\vquestion{runtime speedup}{Poor multithreading performance in .Net} 
and \vquestion{data scraping}{Using multithreading to speed up web crawler written by beautifulsoup4 and python}.

\coincidence{}
Pinto \etal \cite{Pinto:15} observe that 
they 
``did not find questions that ask for advices on
how to use concurrent programming constructs to improve
application performance, which is surprising, since performance is one of the most important motivations for the use
of concurrency and parallelism''. 
In contrast, according to \figref{fig-topics-questions}, our \topic{performance} topic 
includes more than \percent{7} of all of our concurrency questions.  
In addition, our \topic{performance} topic overlaps with Bagherzadeh and
Khatchadourian's \cite{Bagherzadeh:19} big data \topic{performance} topic.

\subsubsection{GUI}
Graphical user interface (GUI) allows for the graphical interaction between a software and its users.
\topic{GUI} is the smallest category with regard to the number of questions and includes question with titles like 
\squestion{}{Force GUI update from UI Thread} and \squestion{}{Object synchronization with GUI Controls}.

\coincidence{}
Our concurrency topic \topic{GUI} overlaps with Rosen and Shihab's
\cite{Rosen:16} \otopic{UI} topic for mobile programming.

\subsection{RQ3: Popularity of Concurrency Topics}
\tabref{fig-topic-pop-table} shows the popularities of our concurrency topics.
These popularities are meausured using 3 metrics average number of
views, favorites and score, in step 9 of our analysis.  
\tabref{fig-topic-pop-table} is sorted by the average number of views
for topics and the highest and the lowest values of popularity metrics are in gray.  A topic
with higher number of views, favorites, and score is more popular.

%

According to \tabref{fig-topic-pop-table}, the \topic{thread safety} topic has the highest views, third highest favorites, and second highest score
whereas \topic{client-server concurrency} has the third lowest views and lowest favorites and score.
\input{fig-topic-pop-table}

\finding{\plusplus{findingcounter}}{
\topic{Thread safety} topic, in \topic{correctness} category, and 
\topic{client-server concurrency}, in \topic{programming paradigms}, are among
the most and least popular concurrency topics, respectively.
%
}

\subsection{RQ4: Difficulty of Concurrency Topics}
\tabref{fig-topic-diff-table} shows the difficulty of concurrency topics. These
difficulties are measured using 2 metrics percentage of questions with no
accepted answers and average median time to get an accepted answer, in step
10 of our analysis. \tabref{fig-topic-diff-table} is sorted by the percentage of
 questions with no accepted answers and the highest and the lowest values of difficulty metrics are
 in gray.  A topic with a higher percentage of questions with no accepted
 answers and median time to receive accepted answers is more difficult.


According to
\tabref{fig-topic-diff-table}, the \topic{irreproducible behavior} topic has the second highest percentage of questions with no accepted
answers and highest time to accepted answers whereas \topic{memory consistency} has the lowest percentage of questions with no accepted answers and 
the second lowest time to accepted answers.  
\input{fig-topic-diff-table}

\finding{\plusplus{findingcounter}}{
\topic{Irreproducible behavior} topic, in  \topic{debugging} category, and 
\topic{memory consistency}, in \topic{correctness}, are among the most and least
difficult concurrency topics, respectively.
}

\subsection{RQ5: Sentiment of Concurrency Topics}
\tabref{fig-topic-sent-table} shows the sentiment of concurrency topics.
These sentiments are measured  
using percentages of their positive, neutral, and
negative polarities, in step 11 of our study. 
\tabref{fig-topic-sent-table} is sorted by the positive polarity and the highest and lowest
values of polarities are in gray. 
A topic is neutral if its neutral polarity is higher than its positive and negative polarities. 
The positive and negative topics are defined similarly.
%

According to \tabref{fig-topic-sent-table}, the neutrality of all concurrency topics is higher than their positivity and negativity.
That is, all concurrency topics are neutral in sentiment.  
\topic{Process life cycle management} and \topic{irreproducible behavior} are the most and least neutral topics.
\topic{Data scraping} and  
\topic{process life cycle management} topics are the most and least positive whereas 
\topic{irreproducible behavior} and \topic{process life cycle management}
are the most and least negative.

Interestingly, in sentiments, the  most positive is not the same as the least negative.
For example, \topic{data scraping} is the most positive topic but not the least negative.  
The same is true for the least positive and most negative sentiments.   

\input{fig-topic-sent-table}

\coincidence{}
Guzman \etal \cite{Guzman:14} analyze the emotions of commit logs in 
29 general GitHub projects. Our observation that all concurrency topics are neutral 
is similar to Guzman \etal's observation that ``the average emotion score
of the commit comments for each of the projects tended to neutrality''.
Similarly, Sinha \etal \cite{Sinha:16} analyze the sentiment of 2,251,585 
commit logs in 28,466 general GitHub projects.
Our observation is also similar to Sinha \etal's observation that ``a
majority [\percent{74.74}] of the sentiment in GitHub projects are categorized as neutral''. 
Tourani \etal \cite{Tourani:14} analyze emotions of 595,673 developer and user emails from Apache's Ant and Tomcat mailing lists.
Finally, our observation is similar to  Tourani \etal's observation that
``19.77\% of the sampled emails were positive, 11.27\% negative. [And, 68.95\% were neutral]''. 

\finding{\plusplus{findingcounter}}{All concurrency topics are neutral in sentiment.}

\finding{\plusplus{findingcounter}}{
\topic{Data scraping} and  \topic{process life cycle management} topics are the most and least positive whereas 
\topic{irreproducible behavior} and \topic{process life cycle management}
are the most and least negative. 
}


\subsection{RQ6: Correlations of Popularity, Difficulty, and Sentiment}
According to Tables \ref{fig-topic-pop-table}-\ref{fig-topic-sent-table},
a topic like \topic{client-server concurrency} has lower popularity, higher difficulty, and a more negative sentiment. 
Such anecdotal evidence could suggest an intuition that there may be correlations between
the popularity, difficulty, and sentiment of concurrency topics where a less popular topic is more difficult and more negative. 
To investigate, we calculate 21 correlations between 3 popularity, 2 difficulty, and 3 sentiment metrics of 
our concurrency topics. These correlations are calculated in step 12 of our analysis.


\tabref{fig-correlations-extend} shows the p-values and directions (positive or negative) of these correlations.
According to \tabref{fig-correlations-extend}, there is a statically significant negative correlation with 95\% confidence between 
the popularity 
and difficulty metrics of concurrency topics, except for the average number of favorites popularity metric 
and the hours to accepted answer difficulty metric.  
That is, more popular concurrency topics are usually less difficult to find answers to their questions.
Similarly, there is a  statically significant negative correlation between the popularity metrics and negative sentiment of 
concurrency topics, except for the average number of views popularity metric and the negative sentiment. That is, more popular topics
are usually less negative. 
Note that these correlation 
do not imply causality.

 

\coincidence{}
Pinto \etal \cite{Pinto:14} study 300  questions and 550 answers related to
software energy consumption on \SO and group them into several themes.
Our observation that the more popular concurrency questions are less difficult
to find answers for 
is in constrast to Pinto \etal's observation that their 
\topic{Measurement} theme is not only their most difficult theme but also the
most popular. Our observation is also in contrast to Rosen and Shihab's
\cite{Rosen:16} observation that finds  Barua \etal's
\cite{Barua:14} popular \topic{mobile development} topic to be difficult.
%
Wang \etal \cite{Wang:18} study the relation between time to answer a question with 46 factors of questions, answers, askers, and answerers 
in 55853, 70336, 7134, and 10776 questions on 4 question and answer websites 
Stack Overflow, Mathematics, Super User, and Ask Ubuntu. 
Our observation 
is also similar to Wang \etal's observation that ``slow-answered
[more difficult] questions are usually associated with rarer [less popular] 
tags than fast-answered questions across the four studied websites''.  

\input{fig-correlations-extended}

\finding{\plusplus{findingcounter}}
{
More popular concurrency topics are usually less difficult and less negative.
}

%

\subsection{RQ7: Race Tool Topics}
\figref{fig-tool-usage} shows the race tool topics that concurrency developers
ask questions about on \SO. These topics are determined in step 13 of our analysis. 
According to \figref{fig-tool-usage}, the questions that developers ask about race tools  
can be grouped into
4 topics:
\topic{root cause identification}, \topic{general}, \topic{\configexec} and \topic{alternative use}. 
In addition, the number of questions that developers ask in each topic is not uniform. 
Developers ask the most questions (\percent{68})
about \topic{root cause identification}
and the least (\percent{9}) about \topic{alternative use} of race tools.  
%
%

\finding{\plusplus{findingcounter}}
{
Questions that concurrency developers ask about race tools can be 
grouped into 4 topics \topic{root cause identification}, \topic{general}, \topic{\configexec} and \topic{alternative use}.
}

\finding{\plusplus{findingcounter}}
{
Developers ask the most number of race tool questions about \topic{root cause identification} and 
the least about \topic{alternative use}.
}


\input{fig-tool-usage}

In the following, we discuss race tool topics and present a set of
example questions for each topic.


\subsubsection{Root Cause Identification}
\topic{Root cause identification} is about using the output of  race detection tools to identify
the source of the reported races in the code.
\topic{Root cause identification} could become challenging because of the high
false positive rate, buggy behavior of the tools, and the difficulty in understanding their outputs.
More than two thirds of race tool questions are about \topic{root cause
identification}. This is in line with the general understanding that 
root cause identification is a major step when debugging software systems.  
\topic{Root cause identification} include questions 
like
%
\squestion{36127075}{ThreadSanitizer says my Atomic Inc/Dec has data races, false positive?},
\squestion{11564606}{Why does valgrind drd think pthread\_barrier\_wait is buggy?},
\squestion{19140720}{intepretation of Valgrind output to figure out the location of data race}, and
\squestion{3035313}{Can't figure out where race condition is occuring}.

\coincidence{}
Sadowski and Yi \cite{Sadowski:14} study the use of widely used ThreadSafety and Thread Sanitizer (TSan) concurrency 
tools at Google and identify 
tool usage themes
by interviewing 
7 developers.    
Our observation that developers are concerned about false positives in \topic{root cause identification}
is similar to Sadowski and Yi's observation that ``another theme that
repeatedly emerged was the importance of a low false positive rate''.
%
%

\finding{\plusplus{findingcounter}}
{
More than two thirds of race tool questions are about \topic{root cause identification},
in line with the general understanding that root cause identification is a major debugging step.
}

\subsubsection{General}
\topic{General} topic is about fundamental guarantees, features, and best practices of race tools 
and includes question with titles like 
\squestion{19090583}{Does one run of Helgrind suffice to guarantee that the given multithreaded implementation is data-race free and deadlock-free?},
\squestion{41316700}{What features does threadsanitizer lack that are supported by helgrind, and vice-versa?}, and 
\squestion{2444787}{What threading analysis tools do you recommend?}. 
Interestingly, about 1 in 7 (\percent{13.5}) of questions that developers ask about race tool are \topic{general} questions. 
This means that usage of race tools and even their basics remains challenging for developers.   


\finding{\plusplus{findingcounter}}
{
Near 1 in 7 questions that developers ask about race tools are \topic{general} questions which means that
the usage of race tools and even their basics remains challenging for concurrency developers. 
}

\mycomment{
\finding{\plusplus{findingcounter}}
{
About 1 in 7 race tool questions are \topic{general} questions.
}
}

\subsubsection{Configuration and execution}
\topic{Configuration and execution} is about the proper configuration and
execution of race detection tools for different execution environments (e.g. 
operating systems and frameworks). \topic{Configuration and execution} includes
question 
like \squestion{26997392}{Is there a way to run
helgrind/drd in android?}, 
\squestion{19770250}{valgrind/helgrind gets killed on stress test}, 
and \squestion{10134638}{Valgrind hanging to profile a multi threaded program}. 
About 1 in 10 race tool questions are about \topic{\configexec}.

\coincidence{}
Our \topic{configuration and execution}
topic overlaps with Bagherzadeh and Khatchadourian's \cite{Bagherzadeh:19} big
data \topic{configuration} category.

\mycomment{
\finding{\plusplus{findingcounter}}
{
About 1 in 10 race tool questions are
about \topic{\configexec}.
}
}

\subsubsection{Alternative Use}
\topic{Alternative use} is about using race tools not only to detect races but also
to monitor, trace, and profile concurrent software for its behavior and performance.
\topic{Alternative use} includes questions 
like   
%
\squestion{9014180}{How to monitor each thread behavior of a multithread (pthread) C++ program on Linux?}, 
\squestion{44085032}{Trace non-reproducable bug in c++}, 
\squestion{1963960}{How to measure lock contention?}, and 
\squestion{25288400}{How can I debug what portion of a multi-threaded C++ program is taking excessive time?}.
About 1 in 10 race tool questions are about \topic{alternative use} of these tools. 

\finding{\plusplus{findingcounter}}
{
Developers ask questions about \topic{alternative use} of race tools for purposes such as profiling and performance 
monitoring of their concurrent software.
}

%% file: fig-topics.tex
\begin{table*}
\centering
\captionskipneg
\captionskiptableneg

\caption{Names, categories and top 10 reduced words of our 27 concurrency topics.
}
\label{fig-topics}
\negspace
\begin{scriptsize}
\newcounter{topiccounter}
\setcounter{topiccounter}{0}

\renewcommand\theadalign{tl}

\newcolumntype{H}{>{\setbox0=\hbox\bgroup}c<{\egroup}@{}}

\setlength{\tabcolsep}{0.4mm}

\begin{tabular}{l H l p{2.6cm} p{5.4cm}}
\rowcolor[HTML]{333333} 
{\color[HTML]{FFFFFF} \textit{\textbf{No.}}} 
& 
&  
{\color[HTML]{FFFFFF} \textit{\textbf{Topic name}}} 
& 
{\color[HTML]{FFFFFF} \textit{\textbf{Category}}} 
& 
{\color[HTML]{FFFFFF} \textit{\textbf{Topic words}}}
\\

\plusplus{topiccounter}&&	
\merge{\colorit{black}{basic concepts}}{debugging}& basic concepts &  code question work answer understand read find edit issu make 
\\

\plusplus{topiccounter}&Thread&    \thead{thread life cycle \\ management} & \hierarchy{\threading}{\model} &  thread main creat run start execut background separ join finish  
\\

\plusplus{topiccounter}&Data structures& 	concurrent collections	& 	\shierarchy{}{\correctness} &  list arrai map element collect iter number item kei add  
\\

\plusplus{topiccounter}&Persistence&	entity management	& \shierarchy{}{\persistence}	&  concurr session spring entiti transact actor collect updat model version  
\\

\plusplus{topiccounter}&&	\thead{object-oriented \\ concurrency} & \shierarchy{}{\paradigm} &  object class method instanc creat static access refer variabl synchron  
\\

\plusplus{topiccounter}&&	task parallelism 			& \shierarchy{}{\model} &  task execut async complet run cancel wait asynchron parallel schedul  
\\

\plusplus{topiccounter}&&	thread sharing 				& \hierarchy{\threading}{\model} &  function variabl pass call pointer pthread argument type return global  
\\

\plusplus{topiccounter}&&	\thead{process life cycle \\ management} 	& \hierarchy{\processing}{\model} &  process child parent termin exit fork creat kill share start  
\\

\plusplus{topiccounter}&&	thread scheduling	&  \hierarchy{\threading}{\model}	&  loop time wait stop run start sleep set check finish 
\\

\plusplus{topiccounter}&&	\merge{thread safety}{multithreading}	&  \shierarchy{}{\correctness}	&  thread java safe multipl multi multithread concurr singl implement applic  
\\

\plusplus{topiccounter}&&	runtime speedup		&  \shierarchy{}{\category{performance}}	&  time core cpu run perform memori number system process machin  
\\

\plusplus{topiccounter}&&	REMOVED		&  \shierarchy{}{}	&  context user set sync local chang synchron save work updat  
\\

\plusplus{topiccounter}&& 	web concurrency	&  \shierarchy{}{\paradigm}&  request servic web server applic user app net respons http  
\\

\plusplus{topiccounter}&& 	\thead{database management \\ systems}	 & \shierarchy{}{\persistence} &  databas tabl queri updat row record lock insert sql transact  
\\

\plusplus{topiccounter}&&	\merge{locking}{\merge{synchronization constructs}{thread synchronization}} & \shierarchy{}{\correctness}	&  lock mutex wait condit releas semaphor acquir deadlock synchron resourc  
%
\\

\plusplus{topiccounter}&&	\merge{\colorit{black}{basic concepts}}{debugging}& basic concepts &  implement make gener librari approach oper case requir solut good
\\

\plusplus{topiccounter}&&	thread pool 				& \hierarchy{\threading}{\model} &  worker pool job number work process task creat limit threadpool  
\\

\plusplus{topiccounter}&&	\merge{data scraping}{data management}				&  \shierarchy{}{\performance} &  data time problem load download work page structur solut url  
\\

\plusplus{topiccounter}&&	\merge{unexpected output}{debugging}	& \shierarchy{}{\category{debugging}} &  code program work run output problem print line result function  
\\

\plusplus{topiccounter}&&	\merge{irreproducible behavior}{debugging} 	& \shierarchy{}{\category{debugging}} &  error test code run problem applic issu window compil crash  
\\

\plusplus{topiccounter}&& 	\thead{event-based \\ concurrency} 	& \shierarchy{}{\paradigm} &  event signal handler timer callback call handl fire receiv slot  
\\

\plusplus{topiccounter}&&	memory consistency			&  \shierarchy{}{\correctness} &  read memori write variabl oper atom cach share synchron access
\\

\plusplus{topiccounter}&&	\thead{producer-consumer \\ concurrency}  & \shierarchy{}{\model} &  queue messag consum produc process item buffer block wait empti  
\\

\plusplus{topiccounter}&&	GUI							&  \shierarchy{}{\category {GUI}} &  updat form gui window applic button control user progress click  
\\

\plusplus{topiccounter}&&	parallel computing			& \shierarchy{}{\category{\model}} &  parallel node loop comput calcul openmp result algorithm mpi function  
\\

\plusplus{topiccounter}&&	\colorit{black}{python multiprocessing} 		& \hierarchy{\processing}{\model} &  python script run multiprocess process function modul command parallel php  
\\

\plusplus{topiccounter}&&	file management & \shierarchy{}{\persistence}	 &  file read write log line open stream directori folder text  
\\

\plusplus{topiccounter}&&	mobile concurrency	& 	\shierarchy{}{\paradigm} &  android imag app activ game view updat frame asynctask devic  
\\

\plusplus{topiccounter}&&	\thead{object-oriented \\ concurrency} & \shierarchy{}{\paradigm} &  call method block return code execut synchron function make time
\\

\plusplus{topiccounter}&&	\thead{client-server \\ concurrency}	&  \shierarchy{}{\paradigm} &  server client connect send socket messag receiv data port read 
\\

\end{tabular}
\end{scriptsize}
\negspace
\negspace

\end{table*}

%% file: fig-topic-question.tex
\begin{figure}[h!]

\captionskipneg

\centering
\includegraphics[width=.85\textwidth]{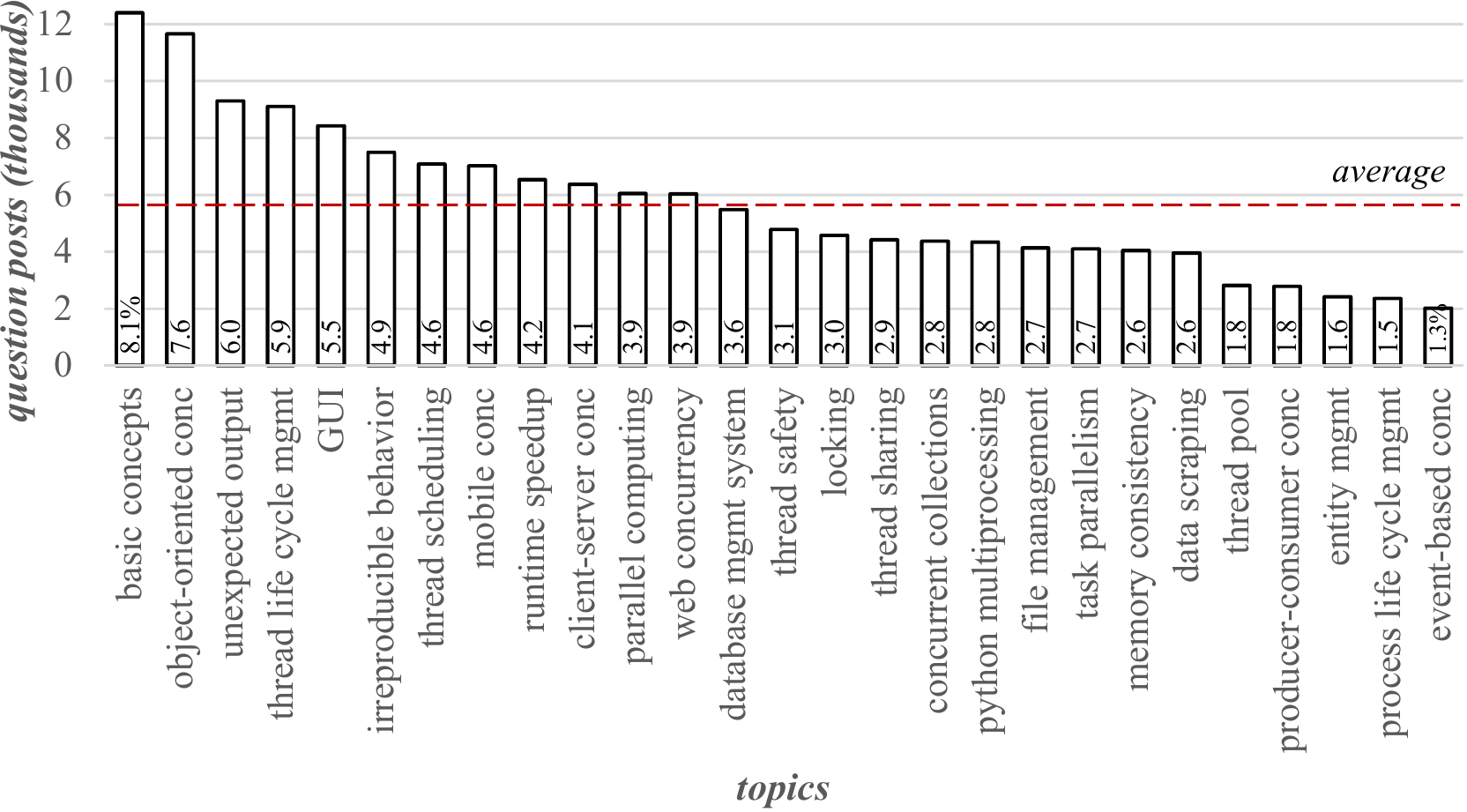}
\negspace
\negspace
\caption{
Concurrency topics and numbers of their questions.
}
\label{fig-topic-question}
\end{figure}

%% file: fig-posts-sunburst.tex
\begin{figure}[h!]
\captionskipneg

\centering
\includegraphics[width=0.85\textwidth]{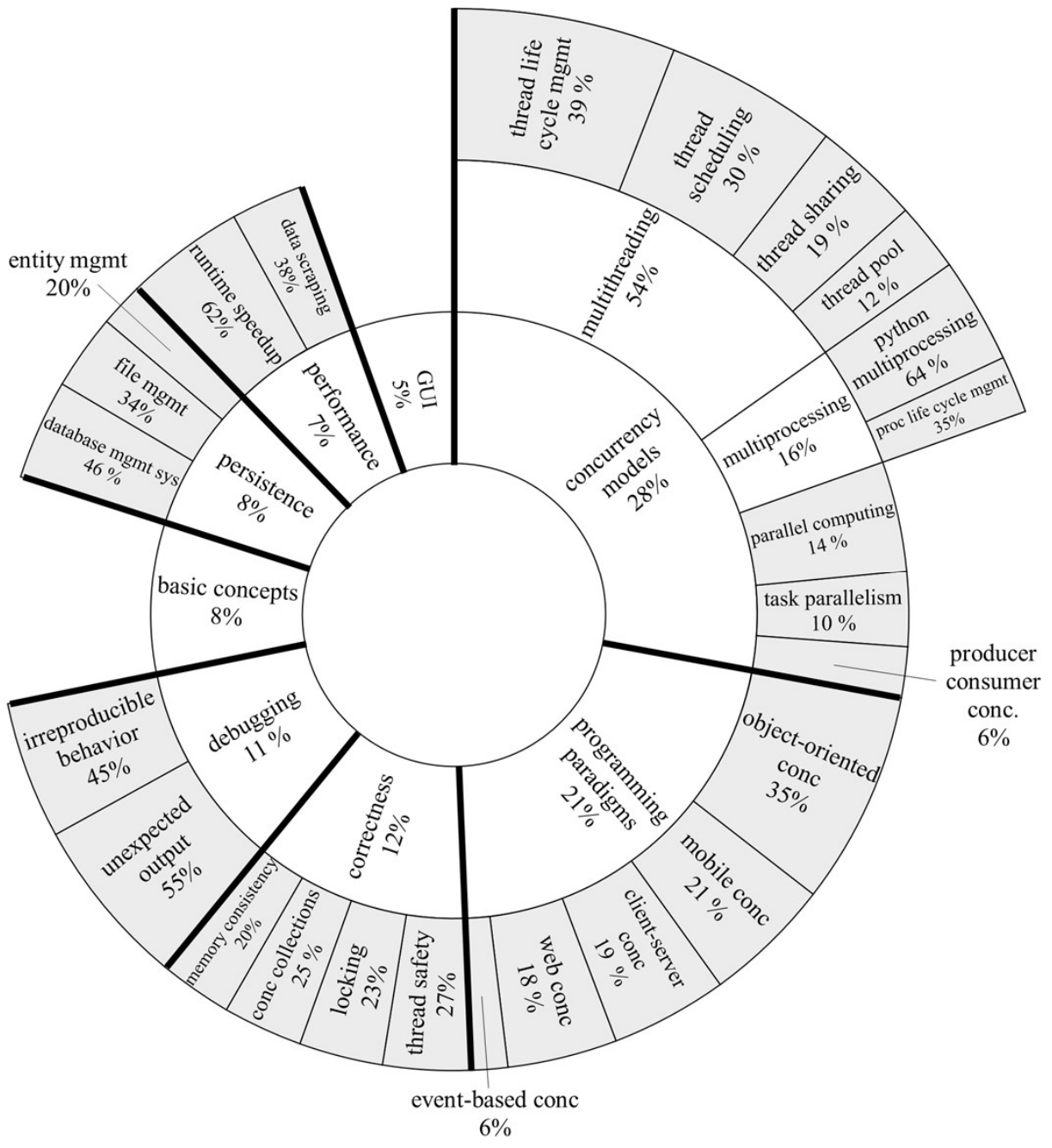}
\negspace
\negspace
\negspace
\caption{
Hierarchy of concurrency topics, in gray, their categories, in white, and percentages of their questions.
}
\label{fig-posts-sunburst}
\end{figure}

%% file: fig-topic-pop-table.tex
\begin{table}[H]
\centering
\captionskipneg
\captionskiptableneg
\caption{Popularity of concurrency topics. The highest and lowest values of popularity metrics are in
gray.}
\label{fig-topic-pop-table}
\negspace
\begin{small}

\setlength{\tabcolsep}{0.66mm}

\begin{tabular}{l c c c}
\rowcolor[HTML]{333333} 
{\color[HTML]{FFFFFF} \textit{\textbf{Topic}}} & {\color[HTML]{FFFFFF} \textit{\textbf{Avg. views}}} &  {\color[HTML]{FFFFFF} \textit{\textbf{Avg. favorites}}} & {\color[HTML]{FFFFFF} \textit{\textbf{Avg. score}}} \\

thread safety & \cellcolor[HTML]{EFEFEF} \nobi{2848} & 1.5 & \cellcolor[HTML]{EFEFEF} \nobi{4.6}
\\ 
basic concepts & 2222 & 1.6 & 4.3
\\ 
task parallelism & 2216 & 1.3 & 4.0
\\ 
locking & 2152 & 1.3 & 3.5
\\ 
thread life cycle management & 2130 & 0.7 & 2.4
\\ 
thread scheduling & 2032 & 0.7 & 2.2
\\ 
process life cycle management & 2004 & 1.0 & 2.6
\\ 
thread pool & 1841 & 0.9 & 2.7
\\ 
object-oriented concurrency & 1773 & 0.8 & 2.6
\\ 
database management systems & 1727 & 0.6 & 1.8
\\ 
thread sharing & 1671 & 0.6 & 2.0
\\ 
GUI & 1664 & 0.5 & 1.6
\\ 
irreproducible behavior & 1647 & 0.6 & 2.3
\\ 
event-based concurrency & 1636 & 0.7 & 2.3
\\ 
python multiprocessing & 1587 & 0.9 & 2.5
\\ 
entity management & 1583 & 0.8 & 2.3
\\ 
memory consistency & 1531 & \cellcolor[HTML]{EFEFEF} \nobi{1.7} & 4.8
\\ 
file management & 1458 & 0.6 & 1.9
\\ 
producer-consumer concurrency & 1311 & 0.8 & 2.2
\\ 
unexpected output & 1304 & 0.5 & 1.6
\\ 
mobile concurrency & 1292 & 0.5 & 1.3
\\ 
runtime speedup & 1276 & 0.9 & 2.7
\\ 
web concurrency & 1252 & 0.8 & 1.9
\\ 
concurrent collections & 1155 & 0.5 & 2.0
\\ 
client-server concurrency & 1083 & \cellcolor[HTML]{C0C0C0} 0.4 & \cellcolor[HTML]{C0C0C0} 1.1
\\ 
data scraping & 1003 & 0.6 & 1.4
\\ 
parallel computing & \cellcolor[HTML]{C0C0C0} 899 & 0.6 & 1.9
\\ 
 
\rowcolor[HTML]{333333} 
{\color[HTML]{FFFFFF} \textit{\textbf{Average}}} & {\color[HTML]{FFFFFF} \textit{\textbf{1641}}} &  {\color[HTML]{FFFFFF} \textit{\textbf{0.8}}} & {\color[HTML]{FFFFFF} \textit{\textbf{2.5}}} \\
\end{tabular}
\end{small}
\negspace
\end{table}

%% file: fig-topic-diff-table.tex
\begin{table}[H]
\centering

\captionskipneg
\captionskiptableneg

\caption{Difficulty of concurrency topics.
The highest and lowest values of difficulty metrics are in gray.  
\newline 
}
\label{fig-topic-diff-table}
\negspace
\begin{small}

\newcolumntype{P}[1]{>{\centering\arraybackslash}p{#1}}

\setlength{\tabcolsep}{0mm}

\begin{tabular}{l c c c}

\rowcolor[HTML]{333333} 
\rule{0pt}{5pt}{\color[HTML]{FFFFFF}\textit{\textbf{Topic}}}&
\rule{0pt}{5pt}{\color[HTML]{FFFFFF}\textit{\textbf{\% w/o acc. answer}}}&
\rule{0pt}{5pt}{\color[HTML]{FFFFFF}\textit{\textbf{Hrs to acc. answer}}} \\

database management systems & \cellcolor[HTML]{EFEFEF} \nobi{51.2} & 1.0 \\ 
irreproducible behavior & 51.1 & \cellcolor[HTML]{EFEFEF} \nobi{2.1} \\ 
web concurrency & 50.7 & 0.9 \\ 
mobile concurrency & 50.4 & 0.8 \\ 
client-server concurrency & 50.4 & 0.9 \\ 
python multiprocessing & 50.3 & 0.9 \\ 
parallel computing & 50.1 & \cellcolor[HTML]{EFEFEF} \nobi{2.1} \\ 
data scraping & 48.9 & 1.0 \\ 
file management & 48.8 & 0.6 \\ 
entity management & 48.0 & 1.8 \\ 
runtime speedup & 48.0 & 0.7 \\ 
thread pool & 47.0 & 0.7 \\ 
process life cycle management & 44.9 & 0.6 \\ 
producer-consumer concurrency & 43.3 & 0.7 \\ 
unexpected output & 41.8 & 0.7 \\ 
GUI & 41.1 & 0.4 \\ 
thread scheduling & 40.8 & 0.4 \\ 
thread life cycle management & 40.4 & \cellcolor[HTML]{C0C0C0} 0.3 \\ 
locking & 40.1 & \cellcolor[HTML]{C0C0C0} 0.3 \\ 
event-based concurrency & 39.7 & 0.6 \\ 
thread safety & 39.6 & \cellcolor[HTML]{C0C0C0} 0.3 \\ 
concurrent collections & 38.6 & 0.4 \\ 
basic concepts & 37.0 & 0.7 \\ 
thread sharing & 35.6 & \cellcolor[HTML]{C0C0C0} 0.3 \\ 
task parallelism & 35.3 & 0.4 \\ 
object-oriented concurrency & 35.2 & \cellcolor[HTML]{C0C0C0} 0.3 \\ 
memory consistency & \cellcolor[HTML]{C0C0C0} 33.2 & 0.4 \\ 
\rowcolor[HTML]{333333} 
\rule{0pt}{5pt} {\color[HTML]{FFFFFF} \textit{\textbf{Average}}} & 
\rule{0pt}{5pt} {\color[HTML]{FFFFFF} \textit{\textbf{43.8}}} & 
\rule{0pt}{5pt} {\color[HTML]{FFFFFF} \textit{\textbf{0.7}}} \\
\end{tabular}

\end{small}
\negspace

\end{table}

%% file: fig-topic-sent-table.tex
\begin{table}[h!]
\centering
\captionskipneg
\captionskiptableneg
\caption{Sentiment of concurrency topics. The highest and lowest sentiment polarities are in gray. 
}
\label{fig-topic-sent-table}
\negspace
\begin{small}

\setlength{\tabcolsep}{0.66mm}

\begin{tabular}{l c c c}
\rowcolor[HTML]{333333} 
{\color[HTML]{FFFFFF} \textit{\textbf{Topic}}} & {\color[HTML]{FFFFFF} \textit{\textbf{Positive\%}}} &  {\color[HTML]{FFFFFF} \textit{\textbf{Neutral\%}}} & {\color[HTML]{FFFFFF} \textit{\textbf{Negative\%}}} \\

data scraping	&	\cellcolor[HTML]{EFEFEF} \nobi{26.7}	&	45.5	&	27.7	\\
database management systems	&	21.7	&	56.8	&	21.4	\\
entity management	&	21.5	&	58.2	&	20.1	\\
thread safety	&	20.4	&	60.1	&	19.3	\\
concurrent collections	&	20.3	&	62.8	&	16.8	\\
mobile concurrency	&	19.4	&	49.4	&	31.1	\\
basic concepts	&	19.3	&	59.7	&	20.9	\\
parallel computing	&	18.8	&	60.3	&	20.8	\\
unexpected output	&	18.3	&	44.7	&	36.9	\\
web concurrency	&	18.2	&	59.1	&	22.6	\\
python multiprocessing	&	17.3	&	55.6	&	26.9	\\
client-server concurrency	&	16.9	&	52.1	&	30.8	\\
GUI	&	16.7	&	52.4	&	30.7	\\
runtime speedup	&	16.6	&	62.4	&	20.9	\\
object-oriented concurrency	&	16.4	&	63.4	&	20.1	\\
irreproducible behavior	&	16.3	&	\cellcolor[HTML]{C0C0C0} 40.6	&	\cellcolor[HTML]{EFEFEF} \nobi{43.0}	\\
memory consistency	&	15.0	&	68.1	&	16.8	\\
producer-consumer concurrency	&	14.4	&	62.4	&	23.1	\\
thread scheduling	&	14.3	&	57.1	&	28.4	\\
task parallelism	&	13.5	&	66.3	&	20.0	\\
file management	&	12.1	&	69.3	&	18.5	\\
event-based concurrency	&	11.7	&	67.9	&	20.2	\\
thread pool	&	11.6	&	71.7	&	16.6	\\
thread life cycle management	&	10.7	&	66.5	&	22.6	\\
thread sharing	&	10.5	&	64.9	&	24.5	\\
locking	&	9.6	&	71.6	&	18.6	\\
process life cycle management	&	\cellcolor[HTML]{C0C0C0} 7.6	&	\cellcolor[HTML]{EFEFEF} \nobi{78.2} &	\cellcolor[HTML]{C0C0C0} 14.0\\

 
\rowcolor[HTML]{333333} 
{\color[HTML]{FFFFFF} \textit{\textbf{Average}}} & {\color[HTML]{FFFFFF} \textit{\textbf{16.4}}} &  {\color[HTML]{FFFFFF} \textit{\textbf{59.8}}} & {\color[HTML]{FFFFFF} \textit{\textbf{24.0}}} \\
\end{tabular}
\end{small}
\negspace
\end{table}

%% file: fig-correlations-extended.tex
\begin{table}[H]
\centering
\captionskipneg
\captionskiptableneg
\caption{
Correlations of popularity, difficulty, and sentiment of concurrency topics. 
The statically significant correlations are in gray.}
\label{fig-correlations-extend}
\negspace
\begin{small}
\setlength{\tabcolsep}{0.66mm}

\begin{tabular}{ccccc}
\multicolumn{2}{c}{}                                                                                                                                                                                    & \multicolumn{3}{c}{\cellcolor[HTML]{333333}{\color[HTML]{FFFFFF} \textit{\textbf{Popularity}}}}                                                                                                                                                                                     \\
\multicolumn{2}{c}{\multirow{-2}{*}{\textit{direction/p-value}}}                                                                                                                                                  & \cellcolor[HTML]{333333}{\color[HTML]{FFFFFF} \textit{\textbf{Avg. views}}} & \cellcolor[HTML]{333333}{\color[HTML]{FFFFFF} \textit{\textbf{Avg. favorites}}}                     & \cellcolor[HTML]{333333}{\color[HTML]{FFFFFF} \textit{\textbf{Avg. score}}}                     \\
\rowcolor[HTML]{EFEFEF} 
\cellcolor[HTML]{333333}{\color[HTML]{FFFFFF} }                                               & \cellcolor[HTML]{333333}{\color[HTML]{FFFFFF} \textit{\textbf{\% w/o acc. answer}}}                     & \signed{-}{}{0.0044}                                                                      & \signed{-}{}{0.01166}                                                                                             & \signed{-}{}{0.001073}                                                                                        \\
\multirow{-2}{*}{\cellcolor[HTML]{333333}{\color[HTML]{FFFFFF} \textit{\textbf{Difficulty}}}} & \cellcolor[HTML]{333333}{\color[HTML]{FFFFFF} \textit{\textbf{Hrs to acc. answer}}}                     & \cellcolor[HTML]{EFEFEF}\signed{-}{}{0.01404}                               & \cellcolor[HTML]{FFFFFF}\signed{-}{}{0.09196}                                                       & \cellcolor[HTML]{EFEFEF}\signed{-}{}{0.02726}                                                                 \\
\multicolumn{1}{l}{}                                                                          & \multicolumn{1}{l}{}                                                                                    & \multicolumn{1}{l}{}                                                        & \multicolumn{1}{l}{}                                                                                & \multicolumn{1}{l}{}                                                                            \\
\multicolumn{2}{c}{}                                                                                                                                                                                    & \multicolumn{3}{c}{\cellcolor[HTML]{333333}{\color[HTML]{FFFFFF} \textit{\textbf{Sentiment}}}}                                                                                                                                                                                      \\
\multicolumn{2}{c}{\multirow{-2}{*}{\textit{direction/p-value}}}                                                                                                                                                  & \cellcolor[HTML]{333333}{\color[HTML]{FFFFFF} \textit{\textbf{Positive}}}   & \cellcolor[HTML]{333333}{\color[HTML]{FFFFFF} \textit{\textbf{Neutral}}}                            & \cellcolor[HTML]{333333}{\color[HTML]{FFFFFF} \textit{\textbf{Negative}}}                       \\
\rowcolor[HTML]{FFFFFF} 
\cellcolor[HTML]{333333}{\color[HTML]{FFFFFF} }                                               & \multicolumn{1}{l}{\cellcolor[HTML]{333333}{\color[HTML]{FFFFFF} \textit{\textbf{\% w/o acc. answer}}}} & \signed{+}{0.16857}{0.21851}                                                & \signed{-}{0.19456}{0.15604}                                                                        & \signed{+}{0.07736}{0.57319}                                                                                         \\
\multirow{-2}{*}{\cellcolor[HTML]{333333}{\color[HTML]{FFFFFF} \textit{\textbf{Difficulty}}}} & \multicolumn{1}{l}{\cellcolor[HTML]{333333}{\color[HTML]{FFFFFF} \textit{\textbf{Hrs to acc. answer}}}} & \cellcolor[HTML]{EFEFEF}\signed{+}{0.34092}{0.01700}                        & \cellcolor[HTML]{FFFFFF}\signed{-}{0.27192}{0.05727}                                                & \cellcolor[HTML]{FFFFFF}\signed{+}{0.06051}{0.67260}                                                                 \\
\multicolumn{1}{l}{}                                                                          & \multicolumn{1}{l}{}                                                                                    & \multicolumn{1}{l}{}                                                        & \multicolumn{1}{l}{}                                                                                & \multicolumn{1}{l}{}                                                                            \\
\multicolumn{2}{c}{}                                                                                                                                                                                    & \multicolumn{3}{c}{\cellcolor[HTML]{333333}{\color[HTML]{FFFFFF} \textit{\textbf{Popularity}}}}                                                                                                                                                                                     \\
\multicolumn{2}{c}{\multirow{-2}{*}{\textit{direction/p-value}}}                                                                                                                                                  & \cellcolor[HTML]{333333}{\color[HTML]{FFFFFF} \textit{\textbf{Avg. views}}} & \multicolumn{1}{l}{\cellcolor[HTML]{333333}{\color[HTML]{FFFFFF} \textit{\textbf{Avg. favorites}}}} & \multicolumn{1}{l}{\cellcolor[HTML]{333333}{\color[HTML]{FFFFFF} \textit{\textbf{Avg. score}}}} \\
\cellcolor[HTML]{333333}{\color[HTML]{FFFFFF} }                                               & \cellcolor[HTML]{333333}{\color[HTML]{FFFFFF} \textit{\textbf{Positive}}}                               & \cellcolor[HTML]{EFEFEF}\signed{-}{0.29915}{0.02860}                        & \cellcolor[HTML]{FFFFFF}\signed{-}{0.15289}{0.28198}                                                & \cellcolor[HTML]{FFFFFF}\signed{-}{0.17947}{0.19498}                                                                 \\
\cellcolor[HTML]{333333}{\color[HTML]{FFFFFF} }                                               & \cellcolor[HTML]{333333}{\color[HTML]{FFFFFF} \textit{\textbf{Neutral}}}                                & \cellcolor[HTML]{FFFFFF}\signed{+}{0.15977}{0.24294}                        & \cellcolor[HTML]{EFEFEF}\signed{+}{0.40229}{0.00469}                                                & \cellcolor[HTML]{EFEFEF}\signed{+}{0.37685}{0.00657}                                                                 \\
\multirow{-3}{*}{\cellcolor[HTML]{333333}{\color[HTML]{FFFFFF} \textit{\textbf{Sentiment}}}}  & \cellcolor[HTML]{333333}{\color[HTML]{FFFFFF} \textit{\textbf{Negative}}}                               & \cellcolor[HTML]{FFFFFF}\signed{-}{0.03714}{0.78629}                                             & \cellcolor[HTML]{EFEFEF}\signed{-}{0.33372}{0.01915}                                                & \cellcolor[HTML]{EFEFEF}\signed{-}{0.28849}{0.04043}                                                                
\end{tabular}

\end{small}
\end{table}

%% file: fig-tool-usage.tex
\begin{figure}[H]
\captionskipneg
\centering
\includegraphics[width=0.65\textwidth]{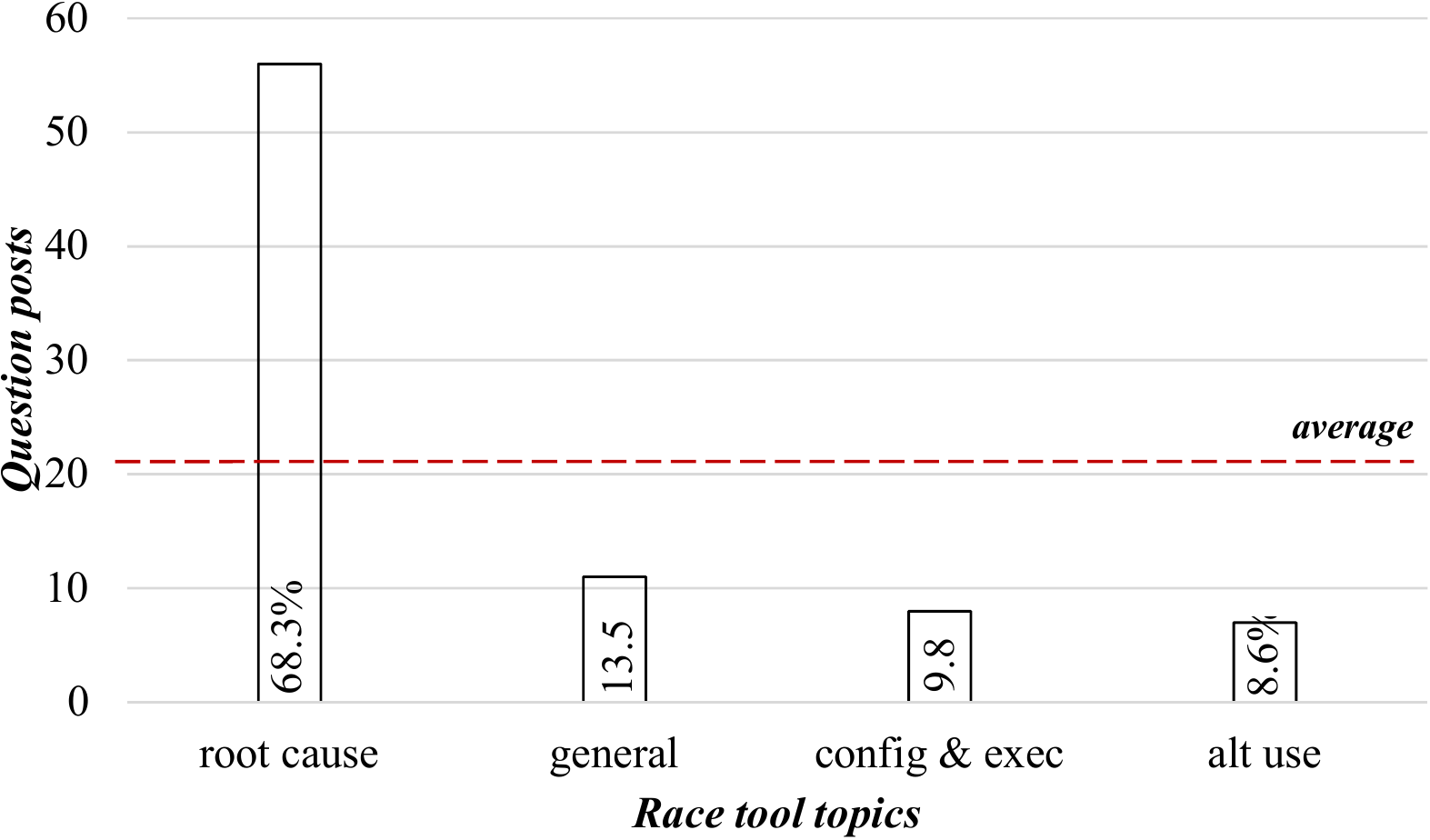}
\negspace
\negspace
\caption{Race tool topics and numbers of their questions.
}
\label{fig-tool-usage}
\end{figure}

%% file: discussion.tex
\section{Implications}
\label{sec:implications}

The results of our study can not only help concurrency developers
and their practice but also the research and education of concurrent
software development to better decide where to focus their efforts.
Members of these communities could use tradeoffs between our concurrency and tool usage
topics as one of several factors they use in their decision making
process. One way to tradeoff a topic against another is
using their popularities, difficulties, sentiments, or the number of their questions.
Obviously, there are many factors that go into tradeoffs that practitioners, researchers, 
and educators make to decide where to focus their efforts.
However, we believe our findings can contribute to inform and improve these decisions.

\input{fig-bubble}
\input{fig-bubble-senti}

To illustrate, \figref{fig-bubble} shows the difficulty of our top 10 most popular concurrency topics.
For simplicity, we use the average number of views as the popularity metric and
and the percentage of questions without accepted answers as the difficulty metrics of these topics. 
The popularity of topics on the vertical axis increases from bottom to top. 
And the difficulty
of topics on the horizontal axis increases from left to right.
The quadrants halve the ranges for the values of the popularity and difficulty metrics.     
In \figref{fig-bubble}, a topic is represented by a circle where the size of the circle is proportional to the 
number of questions in the topic, according to \figref{fig-topic-question}. 
%
%
Similarly, \figref{fig-bubble-senti} shows the sentiment of our top 10 most popular concurrency topics. 
Again, for simplicitly, we use the percentage of positive sentiment as the metric for the sentiment of topics.
The sentiment of topics on the horizontal axis increases from left to right.   


%

\para{Practice}
According to Figures \ref{fig-bubble} and \ref{fig-bubble-senti}, a developer with interest 
in concurrency may make a tradeoff between
the popularities, difficulties, and sentiments of concurrency topics to decide to start their learning 
from the more popular, less difficult, and more positive \topic{task parallelism} topic
before the less popular, more difficult, and less positive \topic{thread pool} topic.  
Similarly, a manager of a concurrency development team may 
 decide to assign a task related to 
the less difficult and more positive \topic{object-oriented concurrency} topic 
to a less experienced developer 
and a task related to the more difficult and less positive \topic{thread pool} topic
to a more experienced developer \cite{Yang:16}.  
\mycomment{
According to \figref{fig-bubble}, a developer with interest in concurrency may decide to start their learning 
on the more popular and less difficult \topic{task parallelism} topic in the \topic{concurrency models} category before 
the less popular and more difficult 
\topic{database management systems} topic in the \topic{persistence} category.
} 

\para{Research}
According to Figures \ref{fig-bubble} and \ref{fig-bubble-senti}, an experienced concurrency researcher 
may make a tradeoffs between  difficulties, popularities, and sentiments of concurrency topics to
decide to focus their research on the more difficult, less popular, and less positive \topic{process life cycle management}
rather than the less difficult, more popular, and more positive \topic{task parallelism} 
in the hope of making more contributions in a less crowded area.
Conversely, a researcher looking to transition anew to the concurrency research may decide to start their research 
from the more popular, less
difficult, and more positive \topic{thread safety} topic.

\mycomment{
According to \figref{fig-bubble}, an experienced concurrency researcher 
may decide to focus their research 
on the more difficult and less popular \topic{thread pool} topic
rather than the less difficult and more popular \topic{thread life cycle management} in the hope of making more contributions in a less crowded area.
Conversely, a researcher looking to transition anew to the concurrency research may decide to start their research on the more popular and less
difficult \topic{thread safety} topic.
}

\para{Education}
According to Figures \ref{fig-bubble} and \ref{fig-bubble-senti}, a concurrency educator may decide to teach the material related to the less difficult, more 
popular, and more positive \topic{thread safety} topic before the more difficult, less popular, and less positive \topic{thread pool}, 
accounting for  the dependencies between these topics. The educator may also decide to prepare more material and spend more time in both the class 
and lab on the more difficult and less positive \topic{thread scheduling} topic  than the less difficult and more positive \topic{thread safety}.
This decision is further supported by the observation that 
there are more questions about \topic{thread scheduling} 
than \topic{thread safety}. 
In addition, according to \figref{fig-tool-usage}, 
the educator teaching about data race tools may decide to spend more time 
on \llstinline{root cause identification} topic with more number of questions 
than 
\llstinline{configuration and execution} topic for these data race tools with less questions.   
%

\finding{\plusplus{findingcounter}}
{Concurrency practitioners, researchers, and educators can use our findings to make decisions about where to focus their effors.}

\mycomment{
\para{Other tradeoffs}
Similarly, \figref{fig-bubble-senti} shows the sentiment of our top 10 most popular concurrency topics. 
For simplicitly, we use percentage of positive sentiment as the measure for sentiment of topics.
The sentiment of topics on the horizontal axis increases from left to right.   

In addition to 
topic popularity and difficulty, practitioners, researchers, and educators can use
popularity and sentiment as well as difficulty and sentiment to tradeoff topics in their decision making process.
\figref{fig-bubble-senti} shows the tradeoff for our top 10 most popular topics using their popularity and positive sentiment.
}

%
%
%

  
%
%
 
%

%

%

%
 
%

%% file: fig-bubble.tex
\begin{figure}[h!]

\captionskipneg

\centering
\includegraphics[width=1\textwidth]{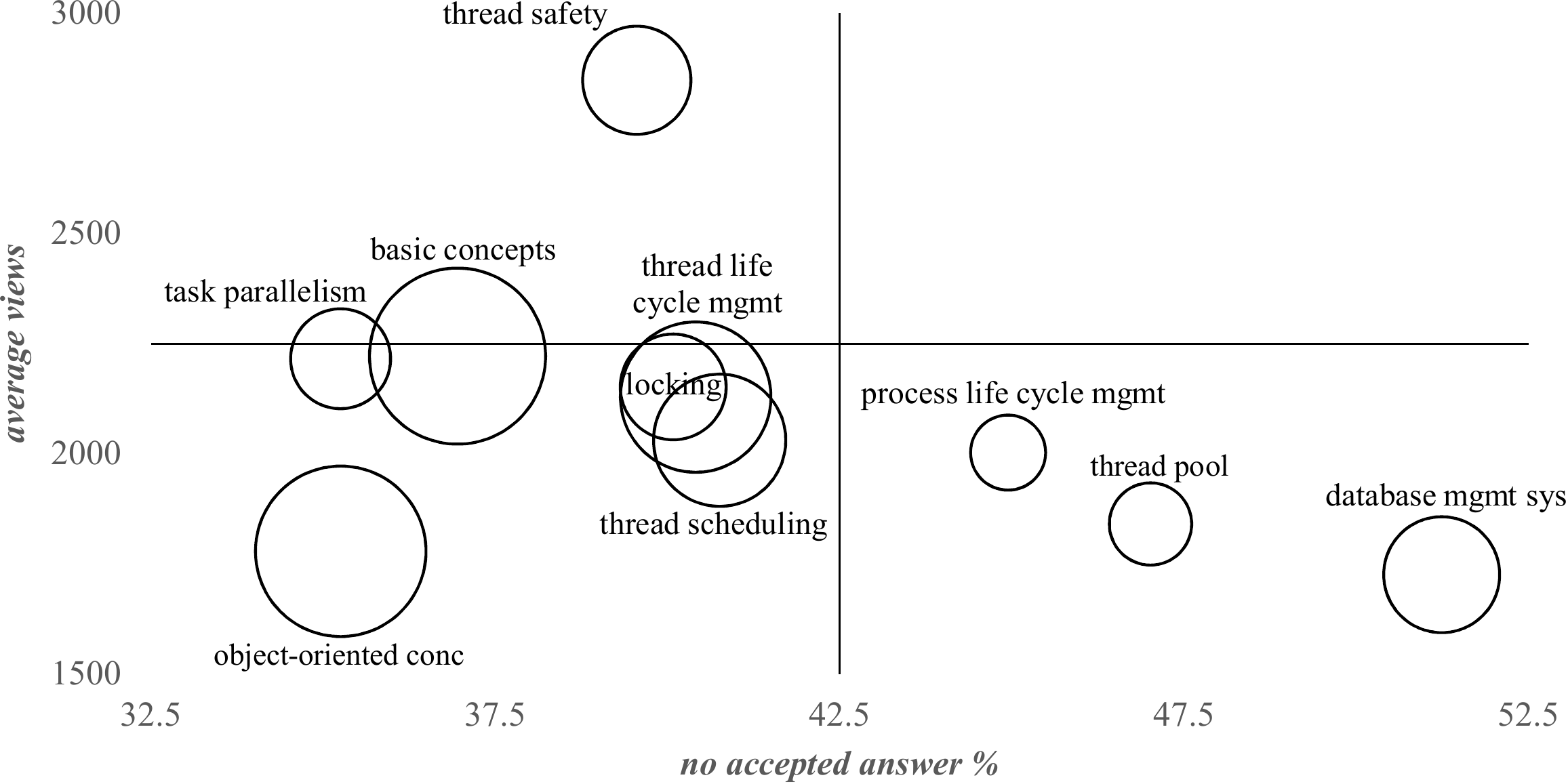}
\negspace
\negspace
\caption{Trading off concurrency topics based on their popularity and difficulty.
}
\label{fig-bubble}
\end{figure}

%% file: fig-bubble-senti.tex
\begin{figure}[h!]

\captionskipneg

\centering
\includegraphics[width=1\textwidth]{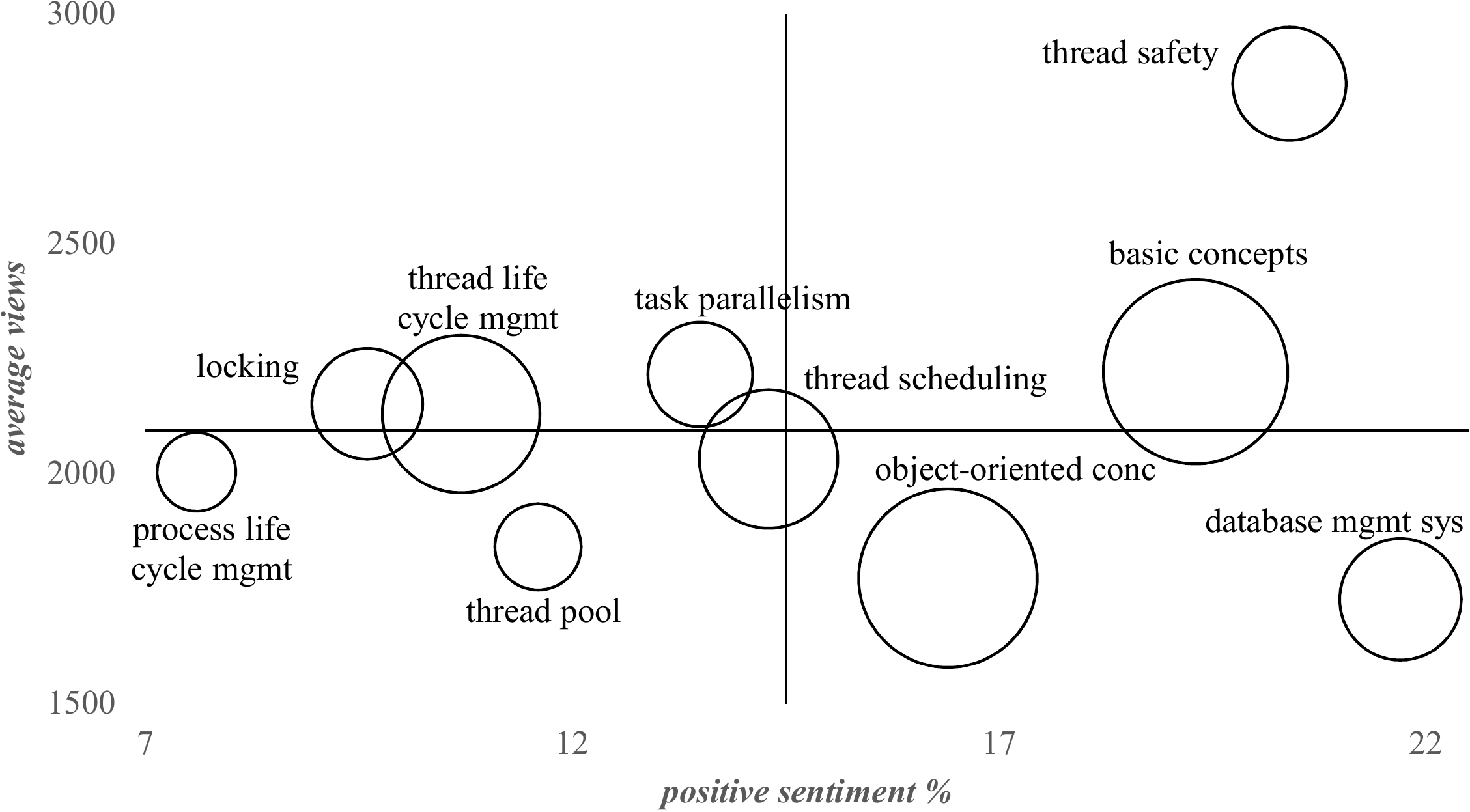}

\negspace
\negspace
\caption{Trading off concurrency topics based on their popularity and sentiment.
}
\label{fig-bubble-senti}
\end{figure}

%% file: validity.tex
\section{Threats to Validity}
\label{sec:threats}


Our use of concurrency tags to identify concurrency questions and answers in \SO  is 
a threat to the validity of our study.
This is because concurrency tags may not be able to identify the complete set of questions and answers related to concurrency.   
To minimize this threat, we use well-known techniques used by previous work \cite{Rosen:16, Yang:16} in 
developing our concurrency tags and perform solid experiments with a broad range 
of tag relevance and significance thresholds for $\sig$ and $\rela$.
Similarly, our use of race tool keywords to identify race tool questions is a threat. 
To minimize this threat, we include industrial race tools that are in use by companies like Google \cite{Sadowski:14} and academic race tools
and techniques surveyed by previous work \cite{Hong:15}. 

%

%

Our use of \SO as the sole dataset 
is 
another threat.
%
 This is because \SO questions and answers may not be  representative of interests, difficulties, sentiments, and race tool usages
of concurrency developers.  
However, the large number of participant developers and questions and answers in \SO along with 
its wide-spread popularity among developers could mitigate this risk.    
Also, unlike some previous work that only use the title of \SO questions in their study 
\cite{Rosen:16}, we use both the title and the body of the questions as well as their 
accepted answers to mitigate this risk.

Our metrics used to measure the popularity, difficulty, and sentiment of topics is another threat.
This is because other factors, such as the 
technology, 
time and day of a question, 
length of the title, body, and code snippets \cite{Wang:18},
of the question can affect its popularity \cite{Treude:11}, difficulty \cite{Wang:18}, and sentiment \cite{Sinha:16,Guzman:14}.   
In fact, Wang \etal \cite{Wang:18} study 46 factors that can affect the time to answer for a \SO question. 
To minimize this threat, we use well-known metrics and tools used by previous work to measure 
popularity \cite{Rosen:16,Yang:16,Bajaj:14}, difficulty \cite{Rosen:16,Yang:16,Treude:11}, 
and sentiment \cite{Calefato:18,Novielli:18,Novielli:18gold,Jongeling:17}.

Our manual labeling of topics is 
%
another threat.
To minimize this threat, we use a well-known approach used by previous work \cite{Bajaj:14} to label topics
using their top 10 words and 15 random questions.  
Determining an optimal value for $K$ when modeling topics is another threat.
To minimize this threat, we use a well-known approach used by  previous work \cite{Barua:14, Bajaj:14, Rosen:16} to find
a reasonable value for $K$ using experiments with a broad range of values for $K$. 
It is well-known that determining an optimal value for $K$ is difficult \cite{Barua:14}.
Parsing \SO dataset, labeling topics from textual contents of questions and answers,
and reduction of words to their 
bases is another threat.
To minimize this threat, 
we use well-known tools used by previous work.
We parse \SO posts using Python elementTree XML API \cite{Rosen:16}, model topics using \Mallet \cite{Barua:14, Bajaj:14, Rosen:16}, and
reduce words 
using the Porter stemming algorithm \cite{Bajaj:14}.

%% file: related.tex
\section{Related Work}
\label{sec:related}

Previous works that are closer to our work study software knowledge repositories, such as \SO, 
to understand
interest \cite{Gyongyi:08,Adamic:08,Hindle:09,Treude:11,Allamanis:13,Barua:14,Rosen:16}, difficulty \cite{Treude:11,Rosen:16,Bajaj:14,Yang:16}, and
sentiment \cite{Garcia:13,Bazelli:13,Guzman:14,Murgia:14,Tourani:14, Ortu:15,Sinha:16,Souza:17} of developers.

\subsection{Concurrency} 
Closest to our work is the work of Pinto \etal \cite{Pinto:15} that uses the 250 most popular concurrency questions on \SO 
to study difficulties that developers face when writing concurrent programs. They categorize these difficulties into a set
of themes  
including theoretical and practical concepts, threading, and first steps themes. 

In other previous work, Pinto \etal \cite{Pinto:15JSS} analyze the code for 2,227 projects to understand the usage of Java's concurrent programming 
constructs and libraries and the evolution of their usage.    
Lin and Dig \cite{Lin:15} study a corpus of 611 widely used Android apps to understand how developers use Android constructs for 
asynchronous concurrency. Blom \etal \cite{Blom:13} study the usage of \lstinline|java.util.concurrent| library in Qualitas corpus.  
Godefroid and Nagappan \cite{Godefroid:08} survey 684 developers to study the spread and popularity of concurrency platforms and models  
at Microsoft. 

In contrast, in this work, we develop a set of concurrency tags to extract concurrency questions and answers from \SO;
group these posts into concurrency topics, categories, and a topic hierarchy;
and analyze popularity, difficulty, and sentiment of these concurrency topics
and their correlations. We also develop a set of race tool keywords to extract
concurrency questions about data race tools and techniques; and group these
questions into race tool topics. Finally, we discuss the implications of our findings for the 
practice, research, and education of concurrent software development,
investigate the relation of our findings with the findings of previous work and
present a sample set of questions for each of our concurrency topics and categories.


\subsection{Non-concurrency}
Rosen and Shihab \cite{Rosen:16} use LDA to model mobile development topics on \SO.
They study popularity and difficulty of their mobile topics and categorize
developers' questions based on platforms for mobile development and type of questions 
that developers ask (why, what, and how).  
Yang \etal \cite{Yang:16} use LDA tuned with a genetic algorithm to model security topics on \SO and
manually organize their topics into 5 categories. They study popularity and difficulty of their security topics. 
Bajaj \etal \cite{Bajaj:14} use LDA to model client-side web development topics using \SO and study
interest of developers in these topics and challenges they face when working with these topics. 
Barua \etal \cite{Barua:14} use LDA to model general topics on \SO.
They study relations of questions and answers of these topics and evolution of 
developers' interest in these topics both in general and for specific technologies.

Gy{\"o}ngyi \etal \cite{Gyongyi:08} and Adamic \etal \cite{Adamic:08} study
Yahoo!Answers posts to determine developers' interests in a set of predefined categories. 
Hindle \etal \cite{Hindle:09} use LDA to model topics related to development tasks 
from commit messages of a standalone 
software project and study evolution of developers' interest in these topics.
Treude \etal \cite{Treude:11} and Allamanis and Sutton \cite{Allamanis:13} study  
\SO questions and answers to infer types of questions that developers ask and determine their difficulties with
these question types.
Bajracharya and Lopes \cite{Bajracharya:12} study logs of Koders, a code search
engine, to learn about general topic of interest to developers in code search.

Guzman \etal \cite{Guzman:14} study the sentiment of commit 
logs and its relation 
with programming languages, team distribution, and approvals 
in 90 \GitHub projects.
They find that commit logs for Java projects tend to be more negative, projects with more
distributed teams tend to have more positive commit logs, and Monday commit logs
are more negative.
Similarly, Sinha \etal \cite{Sinha:16} study the sentiment of 
commit
logs in 
28K \GitHub projects and 
find that a majority of commit logs are neutral,
Tuesday commit logs
are more negative, and a strong correlation exists between sentiment of commit logs 
and the average number of files changed in the commits.

Ortu \etal \cite{Ortu:15} study the sentiment, emotion, and politeness
of issue comments and their relations with the time to fix the
issues in  506K issue comments of 14 projects in the Apache issue tracking
system (JIRA). They find that issues with positive sentiment tend to take less
time to be fixed and there is a weak relation between sentiment, emotion, and
politeness. Souza and Silva \cite{Souza:17} study the   
sentiment of commit logs in 1,262 \GitHub projects and its relation
with builds performed by Travis CI continuous integration server.
They find that commit logs with negative sentiment are more likely to result in broken builds.
Garcia \etal \cite{Garcia:13} study the relation between emotion and activity of contributors in the open source project GENTOO 
using its bug tracking issues and mail archives. They find that contributors with strong positive or negative emotions are more likely
to leave the project. 
Bazelli \etal \cite{Bazelli:13} study the relation between the sentiment of developers' answers on \SO and their reputation and find
that top reputed authors express less negative emotions in their answers. 
Muriga \etal \cite{Murgia:14} study whether software artifacts such as issue reports 
carry emotional information and find that issue reports include emotional information about design choices,
maintenance activity, and colleagues.

In contrast, in this work, we develop a set of concurrency tags to extract concurrency questions and answers from \SO;
group these posts into concurrency topics, categories, and a topic hierarchy;
and analyze popularity, difficulty, and sentiment of these concurrency topics
and their correlations. We also develop a set of race tool keywords to extract
concurrency questions about data race tools and techniques; and group these
questions into race tool topics. Finally, we discuss the implications of our findings for the 
practice, research, and education of concurrent software development,
investigate the relation of our findings with the findings of previous work and
present a sample set of questions for each of our concurrency topics and categories.

%% file: conclusion.tex
\section{Conclusions}
\label{sec:conclusion}
In this work, we conduct a large-scale study on the entirety of \SO to
understand interests, difficulties, sentiments, and tool usages of concurrency
developers.
To conduct this study, we develop a set of concurrency tags to
extract concurrency questions and answers from \SO;
group these questions and answers into concurrency topics, categories, and a
topic hierarchy;
analyze popularities, difficulties, and sentiments of these concurrency topics
and their correlations;
develop a set of tool keywords to extract concurrency questions
about data race tools 
and group these questions into race tool topics; and 
finally  discuss the implications of our findings for the practice, research,
and education of concurrent software development, investigate the relation of our findings
with the findings of the previous work, and present a set of examples questions that developers
ask for each of our concurrency and tool topics.

\mycomment{
In this work, we conduct a large-scale 
study on  the entirety of \SO to understand
interest, difficulty, sentiment, and tool usage of concurrency developers.
We develop a set of concurrency tags to extract concurrency questions and answers from \SO;
group these posts into concurrency topics, categories, and a topic hierarchy;
and analyze popularity, difficulty, and sentiment of these concurrency topics
and their correlations. We also develop a set of race tool keywords to extract
concurrency questions about data race tools and techniques; and group these
questions into race tool topics. 
Finally, we discuss the implications of our findings for the 
practice, research, and education of concurrent software development,
investigate the relation of our findings with the findings of previous work and
present a set of questions for each of our concurrency topics and categories.
}